\documentclass[aps,pra,reprint,superscriptaddress,longbibliography]{revtex4-2}
\usepackage{multirow}
\usepackage{graphicx}
\usepackage{enumerate}
\usepackage{amssymb}
\usepackage{mathtools}
\usepackage{wasysym}

\begin{document}

\def\crta{\vrule height1.41ex depth-1.27ex width0.34em}
\def\dj{d\kern-0.36em\crta}
\def\Crta{\vrule height1ex depth-0.86ex width0.4em}
\def\Dj{D\kern-0.73em\Crta\kern0.33em}
\dimen0=\hsize \dimen1=\hsize \advance\dimen1 by 40pt

\title{Generation of {K}ochen-{S}pecker contextual sets in
  higher dimensions by dimensional upscaling whose complexity
  does not scale with dimension and their applications}

\author{Mladen Pavi\v ci\'c}
\email{mpavicic@irb.hr}

\affiliation{Center of Excellence for Advanced Materials 
and Sensing Devices (CEMS), Photonics and Quantum Optics Unit,\\ 
Ru{\dj}er Bo\v skovi\'c Institute 
and Institute of Physics, 10000 Zagreb, Croatia.}

\author{Mordecai Waegell}
\affiliation{Institute for Quantum Studies, Chapman University,
  Orange, CA 92866, U.S.A.}

\begin{abstract}
  Recently, handling of contextual sets, in particular
  Kochen-Specker (KS) sets, in higher dimensions has been given
  an increasing attention, both theoretically and experimentally.
  However, methods of their generation are diverse, not generally
  applicable in every dimension, and of exponential complexity.
  Therefore, we design a dimensional upscaling method, whose
  complexity does not scale with dimension. As a proof of
  principle we generate manageable-sized KS master sets in up to
  27 dimensional spaces and show that well over 32 dimensions can
  be reached. From these master sets we obtain an ample number of
  smaller KS sets. We discuss three kinds of applications that work
  with KS sets in higher dimensions. We anticipate other
  applications of KS sets for quantum information processing that
  make use of large families of nonisomorphic KS sets.
\end{abstract}

\keywords{quantum contextuality, Kochen-Specker sets, MMP hypergraphs}
 \maketitle

\section{Introduction}
\label{sec:intro}
 
It has been proven that applications in quantum
computation \cite{magic-14,bartlett-nature-14}, quantum steering
\cite{tavakoli-20}, and quantum communication \cite{saha-hor-19}
rely on quantum contextuality and contextual sets. Small contextual
sets, predominantly Kochen-Specker (KS) sets \cite{koch-speck}, in
low dimensional spaces have been implemented in a series of
experiments using photons
\cite{amselem-cabello-09,liu-09,d-ambrosio-cabello-13,ks-exp-03,canas-cabello-8d-14,canas-cabello-14},
neutrons \cite{h-rauch06,b-rauch-09},
trapped ions \cite{k-cabello-blatt-09}, and
solid state molecular nuclear spins  \cite{moussa-09}.
KS sets are sets that prove the KS theorem, i.e., sets of
quantum measurements whose outcomes cannot be predicted by means
of classical (noncontextual) value assignment \cite{cabello-21}.

Some of the applications that have been developed for KS sets,
such as oblivious communication and quantum key distribution,
can make use of any KS set, and having access to a large number of
alternatives may actually increase the security of these protocols
\cite{saha-hor-19,ram-ros-14,horod-10,ram-liu-hor-14,kulikov-17,gallego-10,wehner-08}.
In this paper, we give two applications that work with KS sets
in any dimension, and another for a particular family of KS
sets in even dimensions.

As for implementations of contextual sets, it is
of interest to achieve them in any dimension.
By employing photon orbital angular momentum technique
one can implement “high-dimensional photon quantum gates”
\cite{baba-zeil17,krenn-zeilinger-16} in dimensions up to
eight. Equally so for “multi-photon entanglement in high
dimensions” \cite{malik-zeil2016} or for “time-energy
entanglement with high-dimensional encoding”
\cite{zhong-shapiro-15}. Apparently this is also a
current upper limit for implementation of contextual sets
\cite{canas-cabello-8d-14,budroni-cabello-rmp-22,liu-23}.
However, the long-term goal of a quantum internet, quantum
communication and quantum computation alike, requires
implementations in much higher dimensions than current
experiments can achieve; see \cite{paesani-21} and
\cite{bacco-21}, respectively.

For exploring the scope of future theoretical uses or
experimental implementations of contextuality it would thus
be valuable to develop a universal method of generating
high dimensional contextual sets which may lead to new and
different applications. Existing methods for finding
contextual sets work in high dimensions are diverse and
often limited to a particular family of sets, so the results
are relatively few, and are unevenly distributed. Let us
first review some of these particular cases before moving
on to our methods of generating families of contextual sets.

Ramanathan et al.~\cite{ram-liu-hor-14} consider KS
contextual sets in dimensions $\ge2^{17}$ to obtain as large
violations of non-contextuality inequalities as possible.
Zhan and Hu \cite{zhan-hu-21} consider dimensions up to 20 to
obtain ``dimension-dependent noncontextuality inequalities.''\
Frembs et al.~\cite{frembs-18} show that the contextuality of
qudits of dimension $p^r$, $p$ prime, $r\in\mathbb{N}$, is a
resource for a measurement-based quantum computation.
Waegell and Aravind \cite{waeg-aravind-pra-22} show that
codewords of the binary and ternary Golay codes can be converted
into rays in $\mathbb{RP}^{23}$ and $\mathbb{RP}^{11}$ that provide
proofs of the Kochen-Specker theorem in real state spaces of
dimensions 24 and 12, respectively. Wang et
al.~\cite{wang-sanders-20} make use of $d^n$ dimensional unitary
gate $U\in SU(d^n)$ operating on the $n$-qudit state (where
qudit is of dimension $d$) to provide a ``high-dimensional
quantum computing'' resource and show that a contextual system
solves a problem faster than the classical methods.

Many authors have therefore undertaken an effort to
computationally generate contextual sets for 
possible subsequent usage and application
\cite{planat-saniga-12,waeg-aravind-jpa-15,pavicic-pra-17},
\cite[Supplemental Material]{waeg-aravind-pra-17},
\cite{pm-entropy18,pavicic-entropy-19,pwma-19,pavicic-pra-22,pavicic-quantum-23}. 
They generate KS sets from vectors, projectors, operators, graphs,
stabilizer operations, polytopes, Lie groups, etc. Fortunately,
all these entities can be reduced to hypergraphs. 

So, in order to unify the generation methods and obtain more
general results, exhaustive algorithms for obtaining
contextual hypergraphs from simple vector components, say
$\{0,\pm1\}$, in odd and even dimensional spaces have recently been
proposed \cite{pavicic-pra-22,pm-entropy18}. But, although the
algorithm of this method is valid for any dimension, in practice,
it is limited by its exponential computational complexity. It
faces a computational barrier in dimensions greater than eight.

Therefore, in this paper, we offer a method of generating
manageable-sized master KS sets whose complexity does not grow
with dimension and which can straightforwardly yield smaller sets,
with implementation-optimal vector components $\{0,\pm1\}$, in
higher dimensional spaces. It can provide us with contextual sets
of moderate size in a chosen dimension, each with distinct
structural features that may be relevant for some particular
application. We can generate new sets on demand limited only by
computational resources, and we can also catalog and store many
examples in a database for possible future usage. We stress here
that some contextual sets in higher dimensions are so huge that
most probably the need for very big ones will never materialize,
but smaller sets, whose structural properties can be more easily
understood, may be useful for many applications. Also, ample
contextual sets in a chosen dimension might prove themselves 
indispensable for testing possible new hypotheses made of
a set constructed in the dimension for the purpose. 

The method stems from previous dimensional upscaling approaches
\cite{zimba-penrose,matsuno-07,cabell-est-05,waeg-aravind-pra-17}
which offered several examples as proofs of principle for the
methods, and fleshed out the simplest KS sets in lower dimensions.
We extend and unify these approaches into a method for constructing
KS sets with desired properties in any dimension, which enables us
to find manageable-sized sets in higher dimensions using presently
available computational resources. We give examples in all
dimensions up to 16, then in 27D (27-dim), and finally we give a
blueprint on how one can generate 32D examples. A generation of
non-isomorphic sets in much higher dimensions is just a question
of how much CPU time one is willing to dedicate to these tasks.
The method relies on a remarkable feature of contextual sets that
their ``minimal complexity does not scale with dimension'' as
proved in \cite{waeg-aravind-pra-17}. Our results are discussed
throughout the remainder of this paper.

To describe and handle KS sets we make use of
McKay-Megill-Pavi\v ci\'c-hypergraphs (MMPH)
\cite{pavicic-quantum-23} which we will alternatively simply
call {\em hypergraphs}. In Sec.~\ref{sec:form} we elaborate
on MMPHs in some details. A hypergraph represents each state vector
by a vertex, and states which are mutually orthogonal belong to
{\em hyperedges\/} which we will alternatively simply call
{\em edges\/}. Hypergraphs are the most compact way to represent
a KS set without omitting any structural details, and without
assigning particular state vectors to the vertices (although
some applications may depend in other ways on the particular
assignment of state vectors).

The paper is organized as follows:

In Sec.~\ref{sec:form} we present the hypergraph formalism
we make use of, i.e., the MMPH formalism.

In Sec.~\ref{sec:upscale} we introduce our dimensional
upscaling method.

In Sec.~\ref{sec:review} we present master KS MMPHs we
obtained by means of our dimensional upscaling method
as well as the smaller KS MMPHs we obtained from the
former MMPHs. Subsections \ref{subsec:4}, \ref{subsec:5},
and \ref{subsec:8} provide us with seeds for obtaining
higher dimensional MMPHs in remaining subsections.

In Sec.~\ref{sec:app} we offer three applications of
higher dimensional MMPHs: Larger alphabet (\ref{appA}),
Oblivious communication protocol and communication of
bounded-dimensional systems protocols (\ref{appB}), and
Generalized Hadamard matrices (\ref{appB}).

A discussion is given in Sec.~\ref{sec:disc}.

In Appendix we give KS MMPH's strings
and coordinatizations of masters and the smallest KS MMPHs
for each dimension obtained in Sec.~\ref{sec:review}.

\section{Hypergraph formalism}
\label{sec:form}

An MMPH is a special case of a hypergraph. An $n$D ($n$-dim)
MMPH is a connected {\em hypergraph} $k$-$l$ with $k$ vertices
and $l$ hyperedges (often simply called edges) in which
(i) every vertex belongs to at least one hyperedge;
(ii) every hyperedge contains $n$ vertices;
(iii) no hyperedge shares only one vertex with another hyperedge;
(iv) hyperedges may intersect each other in at most $n-2$ vertices;
(v) graphically, vertices are represented as dots and hyperedges
as (curved) lines passing through them.

We encode MMPHs by means of the printable ASCII characters
for each vertex, with the exception of `space', `0', `+',
`,' and `.'. When all 90 characters are exhausted, we reuse
them prefixed by `{\tt +}' (again for each vertex), when those
are exhausted by `{\tt ++}' and so on. Hyperedges are separated
by `,' and each MMPH is terminated by `.'. There is no limit
on their size. 

A KS MMPH is an $n$D $(n\ge 3)$ MMPH to whose vertices it is
impossible to assign {\rm 1}s and {\rm 0}s in such a way that
the following rules hold: (I) no two vertices in any edge are
both assigned the value $1$; (II) in no edge all of the vertices
are assigned the value 0 \cite[Th.~3.2]{pavicic-quantum-23}.

A given MMPH may or may not have a coordinatization, i.e.,
a representation (of vertices) by means of vectors
in a Hilbert space.  

Our notion of coordinatization differs the notion 
{\em orthonormal representation} used by Lov{\'a}sz
\cite{lovasz-79}. First, Lov{\'a}sz considers an orthonormal
representation of unit vectors in a Euclidean space such
that if $i$ and $j$ are nonadjacent [not connected by an
edge] vertices, then $\mathbf{v}_i$ and $\mathbf{v}_j$ are
orthogonal, while in our notation vectors assigned to adjacent
[connected by an (hyper)edge] vertices are orthogonal.
Second, to Lov{\'a}sz every graph has an orthonormal
representation, while not every MMPH has a coordinatization,
e.g., the 6-3 KS MMP (hyper)graph shown in
\cite[Fig.~1]{pm-entropy18} does not have it.

When a coordinatization is attached to vertices of a KS MMPH,
then the KS theorem \cite{koch-speck,zimba-penrose} states that
such a hypergraph exists. This is in contrast to classical
systems (in, e.g., classical computation) which always
allow the aforementioned assignments of $1$'s and $0$'s.

For a KS MMPH with a coordinatization, its $n$D space becomes
an $n$D Hilbert space spanned by $n$-tuples of mutually
orthogonal vectors, where the $n$-tuples correspond to
hyperedges and vectors to vertices.

A $k$-$l$ MMPH {\em class\/} is a collection of all sub-MMPHs
contained in the  $k$-$l$ MMPH. (A class may contain both
non-KS and KS MMPHs.)

A {\em critical\/} KS MMPH is a KS MMPH which is minimal in
the sense that removing any of its hyperedges turns it into
a non-KS non-contextual MMPH.

A KS MMPH {\em master} is a non-critical KS MMPH which contains
smaller KS proper sub-MMPHs.

The smallest masters contain just one critical proper KS MMPH.
A master may contain non-KS MMPH. A master must contain at
least one proper KS sub-MMPH.

A classical vertex index, $HI_c$, is the number of 1s one can
assign to vertices of an MMPH so as to satisfy the conditions
(I) and (II) of the KS theorem. Maximal (minimal) $HI_c$ is
denoted as $HI_{cM}$ ($HI_{cm}$). It can be proved that
$HI_{cM}=\boldsymbol\alpha$ \cite[Th.~3.2]{pavicic-quantum-23},
where $\boldsymbol\alpha$ is Lov{\'a}sz's independence number 
\cite[p.~192]{gro-lovasz-schr-81}. $HI_{cm}$ enables us to
visually and/or numerically straightforwardly prove any
KS MMPH. See Sec.~\ref{subsec:14}.

\section{Dimensional upscaling method}
\label{sec:upscale}

An essential feature of any
set we consider is that it consists of mutually orthogonal
elements organized in blocks that are themselves mutually linked
so as to form the set. These elements might be operators,
projectors, states, vectors, graph vertices, or hypergraph
vertices. E.g., vector blocks in an $n$-dim space are $n$-tuples
of mutually orthogonal vectors. For all these elements and
blocks we adopt an MMPH representation which we shall often
simply call a hypergraph representation. To arrive at our
{\em master} sets below we make use of vector representation, i.e,
of hypergraphs with coordinatization. To generate smaller sets from
master sets and therefore to form their distributions we make use of
hypergraphs without a coordinatization since then their handling and
computation are much faster and data much more compact. They
reacquire a coordinatizations in a separate step after generation.
    
In \cite{pm-entropy18,pavicic-quantum-23} we generated billions
of KS hypergraphs in dimensions up to eight, directly from simple
vector components. Such a generation in 9+ dimensional
spaces takes too much CPU time even on supercomputers, though.

To generate comparatively small KS hypergraphs in dimensions
higher than those obtained directly
from vector components in \cite{pm-entropy18,pavicic-quantum-23},
we make use of a generalization of methods developed in
\cite{waeg-aravind-pra-17,pavicic-pra-17,pm-entropy18,pavicic-pra-22,pavicic-quantum-23}.

We generalize the Matsuno/Penrose-Zimba method, along with some
other tools, to generate manageable-sized KS hypergraphs which
could then be searched for smaller KS sub-hypergraphs they contain.
Our generalization works by combining two KS sets from lower
dimensions so as to allow the number of unique vertices in the new
combined set to be minimized; it is not guaranteed that the new set
gives a KS hypergraph, so this has to be checked as a separate step.
Given a KS set of dimension $n_1$ with $k_1$ vectors and another of
dimension $n_2 \leq n_1$ with $k_2$ vectors, we can construct a new
set in any dimension $n \leq n_1 + n_2$. The most interesting cases
are with $n < n_1 + n_2$, since then the number of resulting unique
vectors may be significantly less than $k_1 + k_2$.

The method works by extending the vectors of each parent set with
enough vector components 0 to reach dimension $n$. For the first
set, these are all appended at the end of the existing vectors.
For the second set, the $n-n_2$ vector components 0 are
distributed among the first $n_1$ dimensions (the same placements
for every vector in the second set), so that the last $n-n_1$
dimensions are occupied by the nonzero vector components of the
second set, which assures that the new set has nonzero vector
components in all $n$ cardinal directions in the space. At this
point, some of the new vectors from the first set may be identical
to some of the new vectors from the second set, which reduces the
total number of unique vectors in the new set. Thus, to find the
new set with the fewest unique vectors, we consider all
permutations of the locations of the vector components 0 in the
second set, and for each of these, all permutations of the order
of the $n_2$ dimensions.
Because we are using the vector components $\{0,\pm1\}$, this
method allows us to find new sets with $k$ significantly less than
$k_1 + k_2$ for many choices of parent KS sets, and nearly all of
these turn out to give KS hypergraphs---which might not work for
more complicated sets of vector components.

Once we find a minimal set we can check whether it gives a KS
hypergraph or not. If it does, we have a new master set which
likely contains many smaller KS hypergraphs that we can search
for using other methods.

To use the new sets as parents in another round of upscaling to
higher dimension, we want the new KS sets with smallest number of
vectors, so we also check whether any particular vertices can be
removed, along with all of their associated edges, to leave a
smaller KS hypergraph behind, and in many cases this allows us
to remove several extraneous vectors.

Finding new KS sets proceeds in a cyclic way, combining known
KS set in small dimensions to find the new ones in slightly
higher dimensions, keeping the smallest ones, and then repeating
this process and moving to higher dimensions at each iteration.
We probably have not found the smallest sets with vector
components $\{0,\pm 1\}$ for ${\rm dim}>8$, but it seems
likely that we do have them for $4 \leq{\rm dim}\leq 8$. 

Finding the KS sets with the fewest vertices is also important
for computing their set of edges, which takes time that scales
exponentially with both the dimension and the rough number of
vertices in the set.

Our general goal was to find new master KS hypergraphs with a
relatively small number of edges (ideally around 100, but not
more than 1000), which could then be effectively searched for
many smaller KS hypergraphs using our programs
{\textsc{MMPstrip, MMPshuffle}}, and {\textsc{States01}}
\cite{pavicic-quantum-23} with presently available
computational resources. The results of these searches are
presented and discussed in Secs.~\ref{sec:review} and
\ref{sec:disc}.

In particular, this allows us to find the KS hypergraphs with
fewest edges, and the general pattern that emerges is consistent
with the general result that the minimum complexity of KS
hypergraphs does not scale with dimension.  As expected, the
minimum number of vertices seems to grow roughly linearly with
dimension, while the minimum number of edges in $\ge$4D
fluctuates between 9 and 16.  With different vector components,
KS sets with fewer edges are known in certain dimensions (recall
that among the 6D hypergraphs with complex vector components the
smallest one has just 7 edges \cite{lisonek-14}), so the general
minimum is not known in all dimensions and for all vector
components, but this gives an upper bound.

We present and discuss the outcomes of our method in all
dimensions up to 16 and in 27D in Sec.~\ref{sec:review}.

\section{Hypergraphs generated by the upscaling
  method in higher dimensions}
\label{sec:review}

We generate MMPHs with coordinatizations with vector components
from the set $\{0,\pm1\}$. Their figures and distributions
are shown in this section and their strings and coordinatizations
are given in Appendix.

The first three subsections bellow we provide for the sake of
completeness and because we use some of their MMPHs as seeds
for generating MMPHs in higher dimensions via dimensional
upscaling in subsequent subsections. 

\subsection{\label{subsec:3} 3D MMPHs---vector component
   generation}

3D MMPH generated from $\{0,\pm1\}$ is not a KS set
\cite{pavicic-pra-22}.

\subsection{\label{subsec:4} 4D MMPHs---vector component
   generation}

\begin{figure}[ht]
  \center
\includegraphics[width=0.4\textwidth]{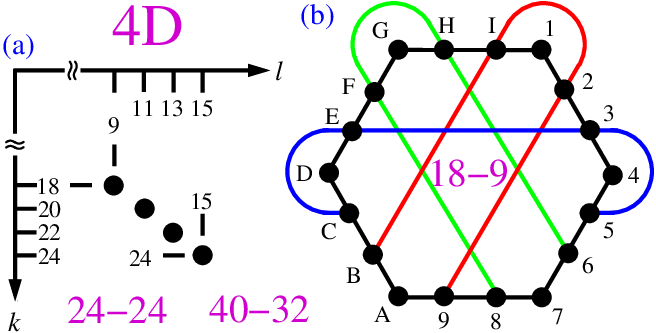}
  \vskip-5pt
\caption{(a) Distribution of the 4D critical KS MMPHs obtained
  from the master 40-32 itself generated by $\{0,\pm 1\}$ vector
  components \cite{pm-entropy18}; 40-32 consists of 2
  MMPHs: a KS 24-24 and a noncontextual 16-8; abscissa is $l$
  (number of hyperedges); (negative) ordinate is $k$ (number of
  vertices); (b) the smallest 4D critical KS MMPH 18-9; $HI_{cM}=
  \boldsymbol\alpha=4$, $HI_{cm}=3$; strings \&\ coordinatizations
  of 24-24, 18-9 are given in Appendix \ref{subsec:4ds}.}
\label{fig:4d}
\end{figure}

We obtain the KS master MMPH 24-24 directly from the vector
component set $\{0,\pm1\}$ via our programs at
\cite{puh-repository}, \textsc{vecfind} (which determines
generates a master from vector components and/or 
whether a vector assignment to vertices in an MMPH is possible),
\textsc{mmpstrip} (which outputs all subsets of the input
MMPHs that have a specified number of hyperedges removed),
\textsc{states01} (which determines whether an MMP diagram
admits a {0,1} (non-dispersive) state), and \textsc{shortd}
(which removes duplicates from input MMPHs):
({\tt vecfind -4d -master -nommp -vgen=0,1,-1 | mmpstrip -U |
grep v24e24| states01 -1 -r 1000 | shortd -G}). From it we
obtain six critical KS MMPHs, shown in Fig.~\ref{fig:4d}(a)
(two 20-11 and two 22-13) within seconds on a single CPU. Also,
via our program {\textsc{mmpsubset}}, we obtain $2^{24}-1$
sub-MMPHs from which we filter out 1232 nonisomorphic KS MMPHs
via {\textsc{states01}} and {\textsc{shortd}}
within minutes on a single CPU \cite{pmm-2-09}. 
Previously, the 24-24 and 18-9 were obtained in \cite{peres}
and \cite{cabell-est-96a}, respectively, by other methods.
See also a diagrammatic representation of 24-24
\cite{waegell-aravind-fp11}. A graphical presentation of 18-9
was first given in \cite[Fig.~3(a)]{pmmm05a} with redundant
cyclically closed hyperedges. In this paper, all MMPHs are 
presented by means of non-redundant graphical presentation of
hyperedges. For instance, in Fig.~\ref{fig:4d}(b) vertices
{\tt C} and {\tt 5} are not connected directly by a straight
line since they are already connected via the {\tt CE34} line.

In the subsequent paragraphs we make use of these three KS MMPHs
as well as of some higher dimensional MMPHs as seeds for generating
MMPHs via dimensional upscaling in higher dimensions.

\subsection{\label{subsec:5} 5D MMPHs---vector
                 component generation}

\begin{figure}[ht]
  \center
\includegraphics[width=0.37\textwidth]{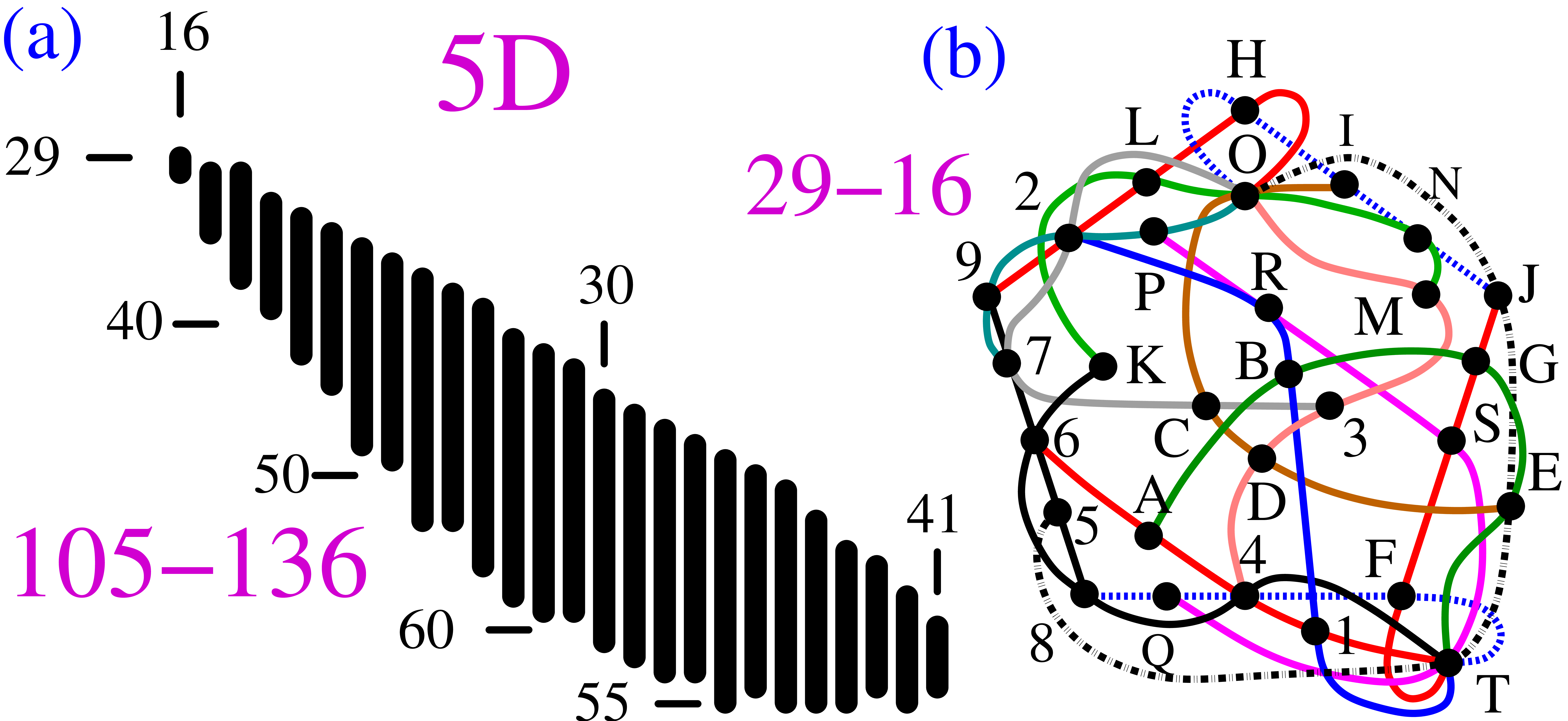}
\caption{(a) Distribution of the 5D critical KS MMPHs obtained
  from the master 105-136 itself generated by $\{0,\pm 1\}$
  vector components \cite{pavicic-pra-22}; abscissa is $l$
  (number of hyperedges); (negative) ordinate is $k$ (number of
  vertices); Cf.~Fig.~\ref{fig:4d}(a);
  (b) One of the two smallest five-dimensional 29-16 critical
  KS MMPH; $HI_{cM}$=$\boldsymbol\alpha$=$7$, $HI_{cm}$=$3$;
  strings and coordinatizations  are given in
  Appendix \ref{subsec:5ds}.}
\label{fig:5d}
\end{figure}

Arbitrarily exhaustive number of critical 5D KS MMPHs shown
in Fig.~\ref{fig:5d}(a) were obtained in
\cite{pwma-19,pavicic-pra-22} directly from the set
$\{0,\pm1\}$ by vector component generation.
Critical 29-16 (Fig.~\ref{fig:5d}(b)) serves us as a seed
for generating some master MMPHs in higher dimensions.

\subsection{\label{subsec:6} 6D MMPHs---dimensional
                 upscaling}

\begin{figure}[ht]
  \center
\includegraphics[width=0.4\textwidth]{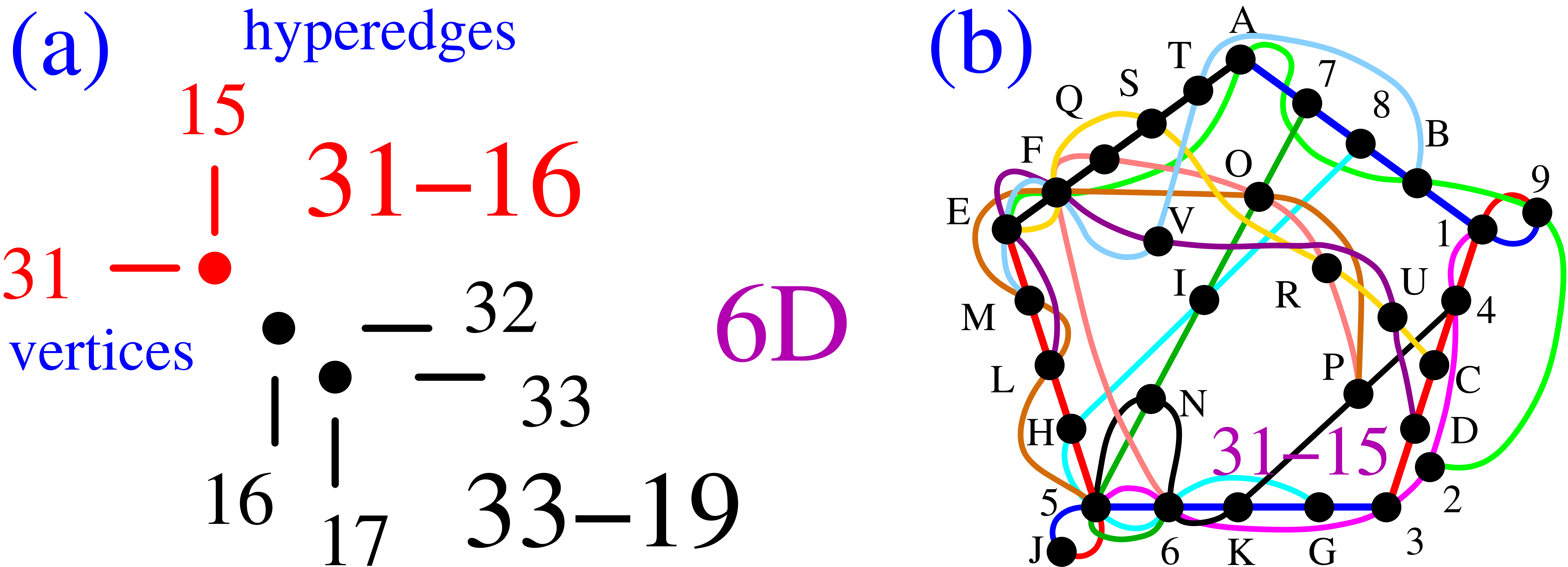}
\caption{(a) Distribution of the smallest 6-dim critical KS
  MMPHs obtained from the masters 31-16 and 33-19 by
  dimensional upscaling; abscissa is $l$ (number of hyperedges);
  (negative) ordinate is $k$ (number of vertices);
  Cf.~Fig.~\ref{fig:4d}(a); (b) the smallest critical
  KS MMPHs 31-15 in the 31-16 class; $HI_{cM}=
  \boldsymbol\alpha=7$, $HI_{cm}=2$; strings and
  coordinatizations of 31-15, 32-16, and 33-17 are given in
  Appendix \ref{subsec:6ds}.}
\label{fig:6d}
\end{figure}

Six dimensional {\em Hilbert spaces} are inhabited by either
spin-$\frac{5}{2}$ systems or by qubit-qutrit systems
 (${\cal H}^6={\cal H}^2\otimes{\cal H}^3$).
The former representation in a complex space has been implemented
in \cite{canas-cabello-14}. It turned out to have a fairly small
master MMPH \cite{pm-entropy18,pwma-19,pm-paris19} and therefore
it had been easy to find all its subgraphs. In the
real space, the 332-1408 master MMPH generated by $\{0,\pm1\}$
components is huge and already at the time of its generation
\cite{pavicic-pra-17,pm-entropy18} we were well aware that 
then obtained 34-16 critical might not be the smallest. Our
dimensional upscaling confirms that conjecture and provides us
with a 31-{\bf 15} critical MMPH shown in Fig.~\ref{fig:6d}(b).
We obtain it from the master 31-16. From another 33-19 master
we obtain three additional MMPHs, one of which (32-16) is also
smaller then the 34-16. Their distributions are shown in
Fig.~\ref{fig:6d}(a).

\subsection{\label{subsec:7} 7D MMPHs---dimensional
                 upscaling}

The vector components $\{0,\pm1\}$ in a 7D spin-3 space
generate an 805-9936 master MMPH which generates a class whose
partial distribution is shown in \cite{pavicic-pra-22}. Its
vectors were automatically generated from the master MMPH by
means of our programs {\textsc{MMPstrip, MMPshuffle}}, and
{\textsc{States01}}.

However, such a direct exhaustive generation  takes too
much CPU time on a supercomputer and therefore we here generate
a partial distribution of KS MMPHs with small number of vertices
and hyperedges (not obtained in \cite{pavicic-pra-22}) from
master 47-176 obtained here via dimensional
upscaling as shown in Fig.~\ref{fig:7d}(a).
\begin{figure}[h]
  \center
\includegraphics[width=0.4\textwidth]{7d-cai2.eps}
\caption{(a) Distribution of a million of 7D critical KS
  MMPHs from the 47-176 master; Cf.~Fig.~\ref{fig:4d}(a);
  (b) the smallest criticals 34-14 from the 47-176 master;
  $HI_{cM}=\boldsymbol\alpha=7$, $HI_{cm}=3$; strings and
  coordinatizations of 33-14 are given in
  Appendix \ref{subsec:7ds}.}
\label{fig:7d}
\end{figure}
The smallest
critical\break 34-{\bf 14} MMPH from this class is shown in
Fig.~\ref{fig:7d}(b). 
Note the cutoff level at the 63 vertices characteristic of all
masters that are not generated directly from the vector
components. The ones that are we sometimes call {\em supermasters\/}
and the former ones a {\em submasters}. Analogous cutoffs are
evident for all higher dimensional sub-masters below.

\subsection{\label{subsec:8} 8D MMPHs---vector
                 component generation}

\begin{figure}[h]
  \center
\includegraphics[width=0.4\textwidth]{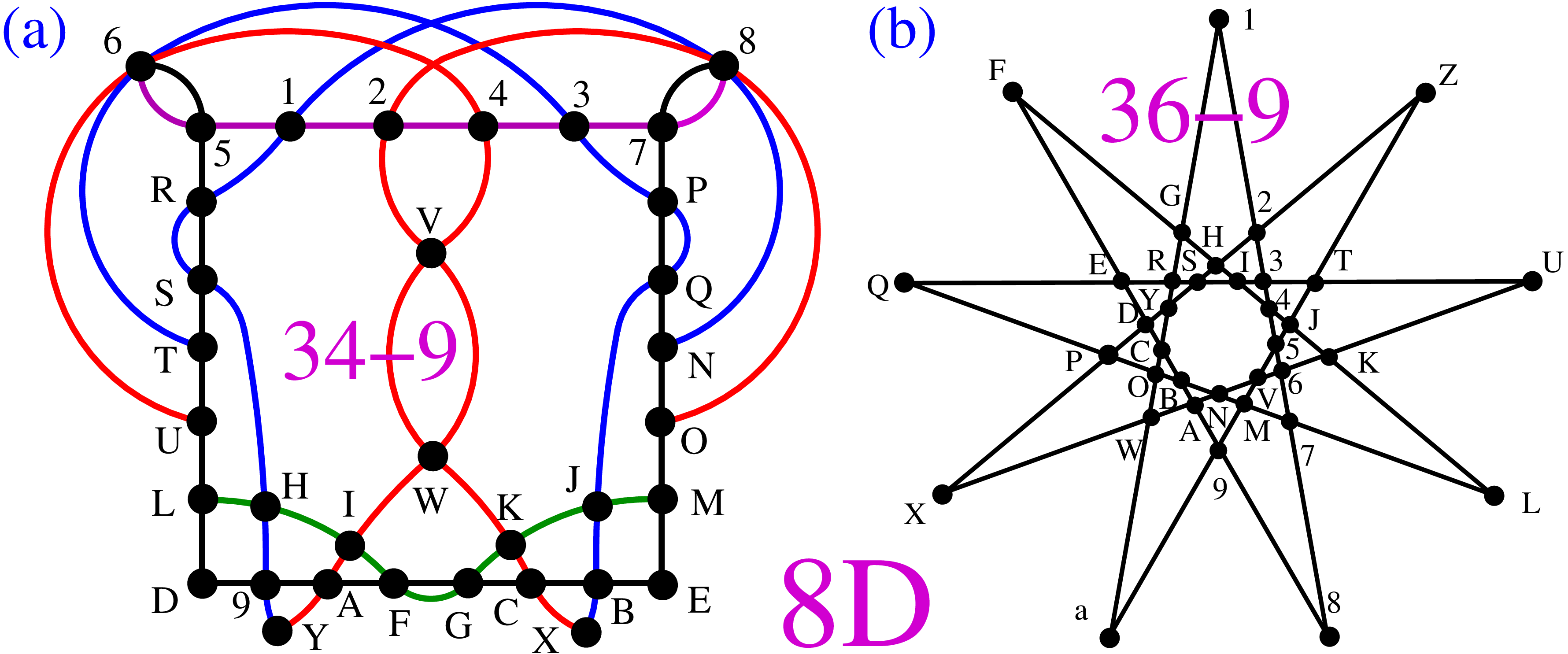}
\caption{(a) one of the two smallest non-isomorphic criticals
  34-9 \cite{pavicic-pra-17}; $HI_{cM}= \boldsymbol\alpha=4$,
  $HI_{cm}=3$; Cf.~Fig.~\ref{fig:4d}(a); (b) star-like 36-9
  \cite{pavicic-pra-17} proved by the $S$-$H$ theorem
  (\cite{lisonek-19}; see Sec.~\ref{appC}); strings and
  coordinatizations are given in Appendix \ref{subsec:8ds}.}
\label{fig:8d}
\end{figure}

Billions of 8D critical KS MMPHs obtainable from the
3280-1361376 $\{0,\pm\}$ master are given in
\cite{pavicic-pra-17}. So, there is no need to employ the
dimensional upscaling to replicate some of them here.
However, we do make use of the smallest 34-{\bf 9} KS MMPH,
shown in Fig.~\ref{fig:8d}(a), as a seed for
dimensional upscaling of MMPHs in higher dimensions.

\subsection{\label{subsec:9} 9D MMPHs---dimensional
                 upscaling}

\begin{figure}[ht]
\center
\includegraphics[width=0.48\textwidth]{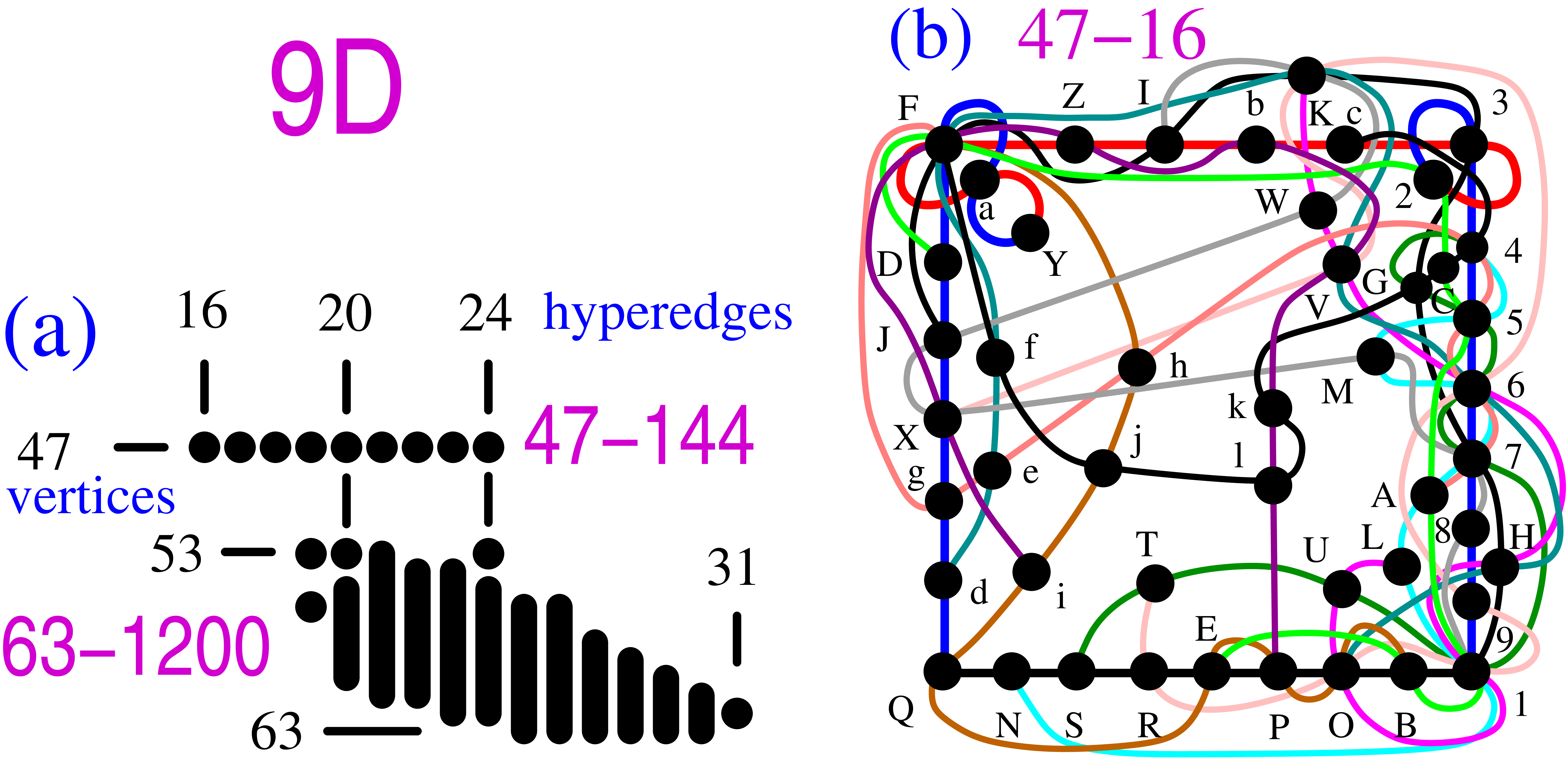}
\caption{(a) Distribution of 41,000 9D critical KS MMPHs 
  obtained via dimensional upscaling from the 47-144 and
  63-1200 master MMPHs; Cf.~Fig.~\ref{fig:4d}(a);
  (b) the smallest critical  47-16
  from the 47-144 class; $HI_{cM}=\boldsymbol\alpha=7$,
  $HI_{cm}=3$; strings and coordinatizations of 47-16
  are given in Appendix \ref{subsec:9ds}.}
  \label{fig:9d}
\end{figure}

Two entangled qutrits live in a 9D space and in
\cite{pavicic-quantum-23} we generated their MMPH supermaster
from $\{0,\pm1\}$ components. It consists of 9,586 vertices and
12,068,704 hyperedges and that is too huge for a direct
generation from the master MMPH. However, small critical KS MMPHs
can be obtained by dimensional upscaling which yields 47-144 and
63-1200 master MMPHs whose distributions are shown in
Fig.~\ref{fig:9d}(a). Critical KS MMPHs are generated
from the masters via {\textsc{States01}}. 

\subsection{\label{subsec:10} 10D MMPHs---dimensional
                 upscaling}

From the spin-$\frac{9}{2}$ MMPH masters 52-141, 60-96, 
and 74-610, whose distributions are shown in 
Fig.~\ref{fig:10d}(a),
\begin{figure}[ht]
\center
\includegraphics[width=0.48\textwidth]{10d-ss.eps}
\caption{(a) Distribution of 10D critical KS
  MMPHs obtained from the 52-141, 60-96, and 74-610 master
  MMPHs; Cf.~Fig.~\ref{fig:4d}(a); (b) the smallest critical
  KS MMPH 50-15 from the 52-141 class; $HI_{cM}=
  \boldsymbol\alpha=7$, $HI_{cm}=2$; (c) star-like KS
  MMPH 55-11 whose existence was proved by the $S$-$H$ theorem
  (\cite{lisonek-19}; see Sec.~\ref{appC}; the string and
  coordinatization of the 50-15 and the string of the 55-11 are
  given in Appendix~\ref{subsec:10ds}; it is a critical KS MMPH
  with a parity proof.}
  \label{fig:10d}
\end{figure}
we obtain critical MMPHs,
the smallest of which (50-{\bf 15}) is shown in
Fig.~\ref{fig:10d}(a). Their strings might have gaps
in characters (e.g., {\tt 3} and {\tt K} are missing in
50-15). The gaps can be closed by program {\textsc{MMPshuffle}}
if needed (e.g., for further processing).

The 10D KS master MMPHs are generated by means of particular
combinations of hypergraphs from smaller dimensions.
See Sec.~\ref{subsec:12}.

\subsection{\label{subsec:11} 11D MMPHs---dimensional
                 upscaling}

\begin{figure}[ht]
\center
\includegraphics[width=0.48\textwidth]{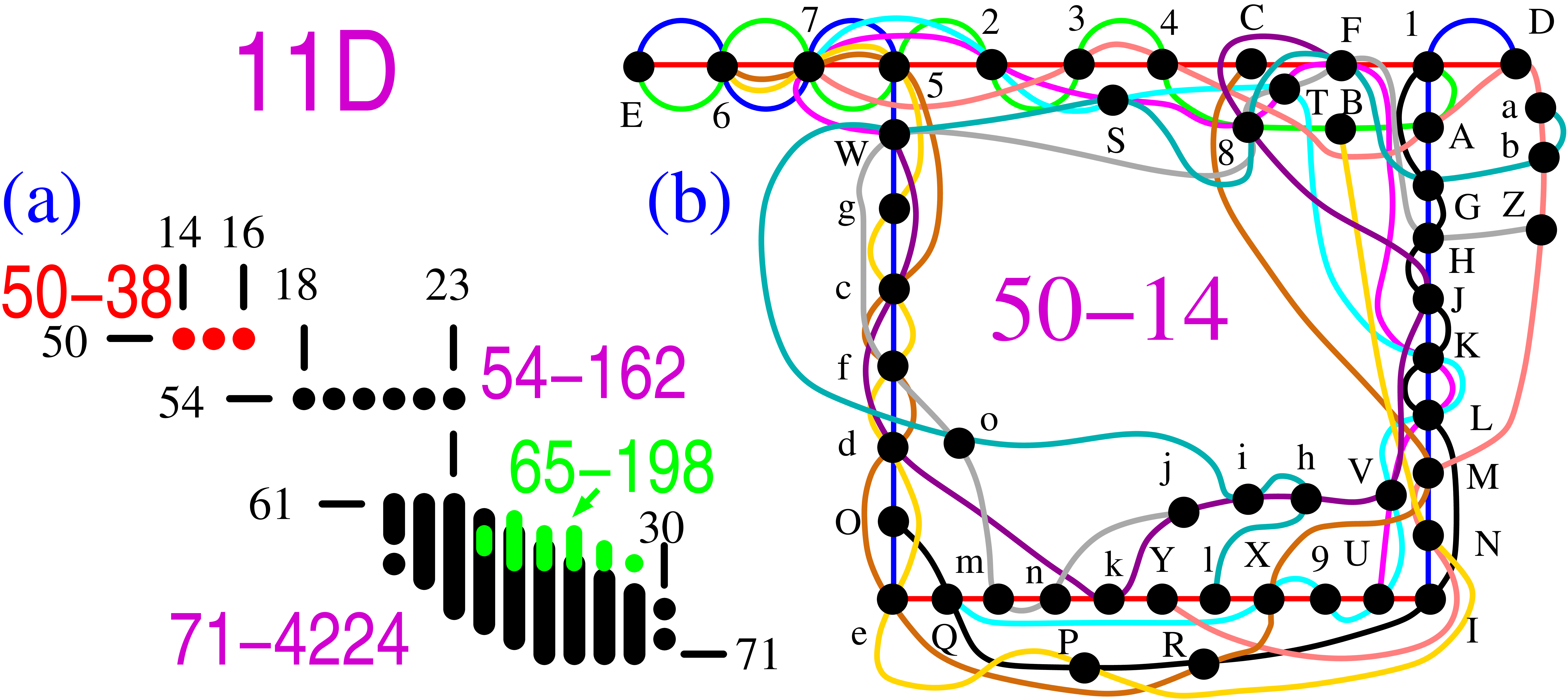}
\caption{(a) Distribution of 11D critical KS MMPHs obtained
  from the 50-38, 54-162, 65-198, and 71-4224 master MMPHs via
  dimensional upscaling; Cf.~Fig.~\ref{fig:4d}(a); (b) the
  smallest critical 50-14 we obtained from the 50-38;
  $HI_{cM}=\boldsymbol\alpha=6$, $HI_{cm}=3$; strings and
  coordinatizations of 50-14 are given in Appendix
  \ref{subsec:11ds}.}
\label{fig:11d}
\end{figure}

Spin-5 11D master MMPHs with $\{0,\pm 1\}$ vector components
generate smaller critical KS MMPHs by the dimensional upscaling.
The master MMPHs are 50-38, 54-162, 65-198, and 71-4224. From them
we generated the following critical KS MMPHs: 50-14, \dots,60-16,
54-18,\dots, 54-23, and 61-21,\dots, 70-30 (each of which
is coming in many non-isomorphic instances) whose distributions
are shown in Fig.~\ref{fig:11d}(a). We give the smallest
one 50-{\bf 14} in Fig.~\ref{fig:11d}(b).

\subsection{\label{subsec:12} 12D MMPHs---dimensional
                 upscaling}

\begin{figure}[ht]
\center
\includegraphics[width=0.49\textwidth]{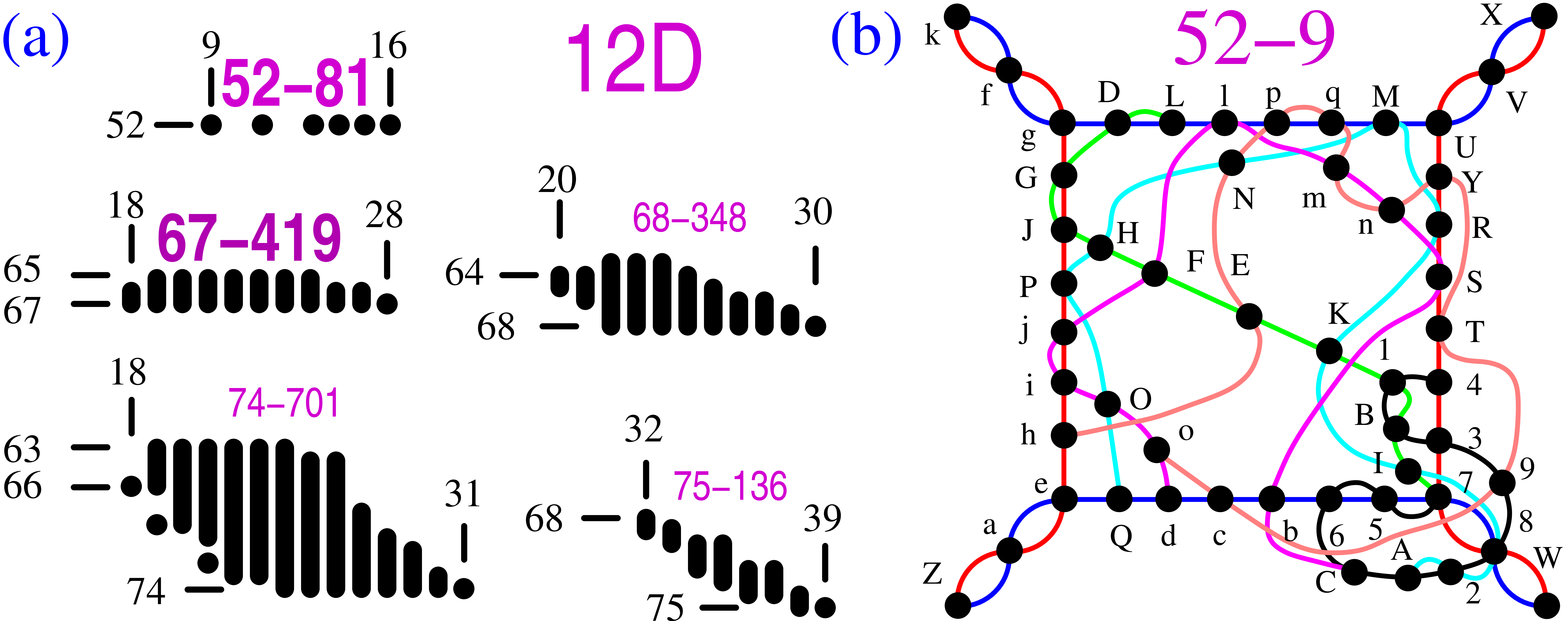}
\caption{(a) Distribution of 12D critical KS MMPHs obtained
  from the 52-81, 67-419, 68-348, 74-701, and 75-136 master MMPHs
  which are themselves obtained via dimensional upscaling;
  Cf.~Fig.~\ref{fig:4d}(a); (b) the smallest critical KS MMPH
  52-9 we obtained from the 52-81 master;
  $HI_{cM}=\boldsymbol\alpha=4$, $HI_{cm}=3$;
  strings and coordinatizations of 52-9 are given in
  Appendix \ref{subsec:12ds}.}
\label{fig:12d}
\end{figure}

12D MMPHs can be represented either by 
spin-$\frac{11}{2}$ systems or by two qubits and a qutrit
(${\cal H}^{12}={\cal H}^2\otimes{\cal H}^2\otimes{\cal H}^3$).
The critical 52-{\bf 9} shown in Fig.~\ref{fig:12d}(b) 
is one
of 502 non-isomorphic critical 52-9s from the 52-81 master. It
represents a partial constructive proof (for MMPHs with
$\{0,\pm1\}$ component coordinatization) of the result that
MMPHs in an even dimensional space $n\ge10$ require at most
nine hyperedges \cite{waeg-aravind-pra-17}. 

The 12D 52-81 master was obtained by combining the 34-9 set in
8D with the 18-9 in 4D. The 67-419 master is obtained as follows.
First, we combine two 18-9 sets in 4D to get a 35-32 in 7D.
Next, we combine the 35-32 in 7D with the 18-9 in 4D to get a
52-141 set in 10D (see Fig.~\ref{fig:10d}(a)). And finally, we combine this 52-141 set
in 10D with the 18-9 in 4D to get a 67-419 in 12D.
Similarly with the remaining four masters. We then apply
\textsc{MMPstrip} and \textsc{States01} programs to them to obtain
distributions of all smaller KS hypergraphs contained in the
masters. The five distributions are shown in the figure. The
smallest KS hypergraph 52-9 (shown in Fig.~\ref{fig:12d}(b)) is
contained in just one of the masters (52-81).

\subsection{\label{subsec:13} 13D MMPHs---dimensional
                 upscaling}

\begin{figure}[ht]
\center
\includegraphics[width=0.48\textwidth]{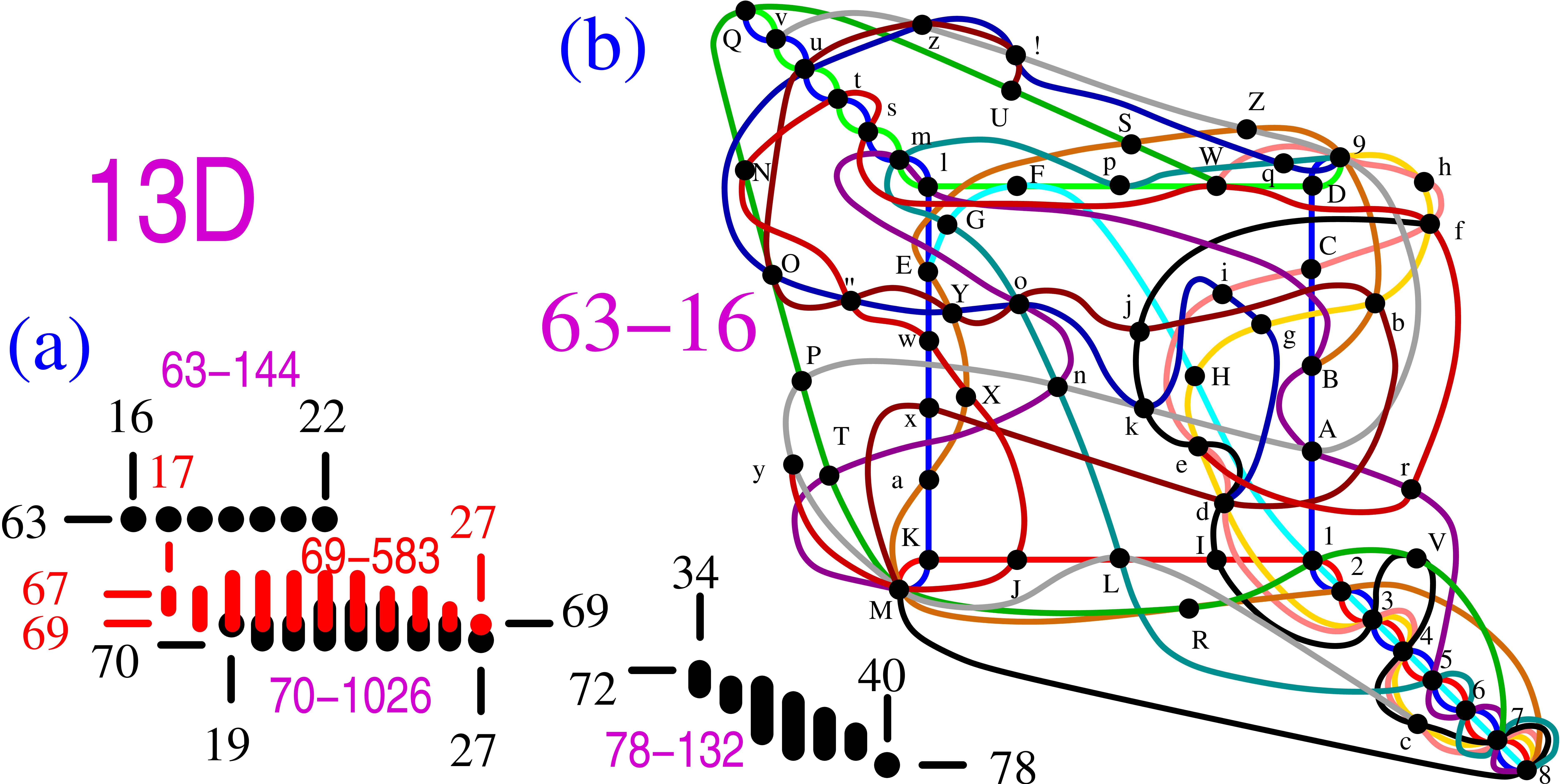}
\caption{(a) Distribution of 100,000 13D critical KS MMPHs
  obtained from the 63-144, 69-583, 70-1026, and 78-132 master
  MMPHs obtained via dimensional upscaling; Cf.~Fig.~\ref{fig:4d}(a);
  (b) one of the smallest criticals, KS MMPH 63-16 we obtained from
  the 63-144 master; $HI_{cM}=\boldsymbol\alpha=7$, $HI_{cm}=3$;
  strings and coordinatizations of 63-16
  are given in Appendix \ref{subsec:13ds}.}
\label{fig:13d}
\end{figure}

Spin-6 13D criticals MMPHs shown in Fig.~\ref{fig:13d} are less
abundant than the ones obtained in lower dimensions. This is due
to the complexity of generation of MMPHs which forced us to
generate as small masters as possible in order to shorten the
runtimes of generation. However, although the number of vertices
and hyperedges is limited we still get a high number of
non-isomorphic criticals. Actually, among 100,000 of generated
MMPHs there are no two isomorphic ones. Their distributions are
shown in Fig.~\ref{fig:13d}(a). The smallest critical
is 63-{\bf 16} (Fig.~\ref{fig:13d}(b)).

\subsection{\label{subsec:14} 14D MMPHs---dimensional
                 upscaling}

\begin{figure}[ht]
\center
\includegraphics[width=0.48\textwidth]{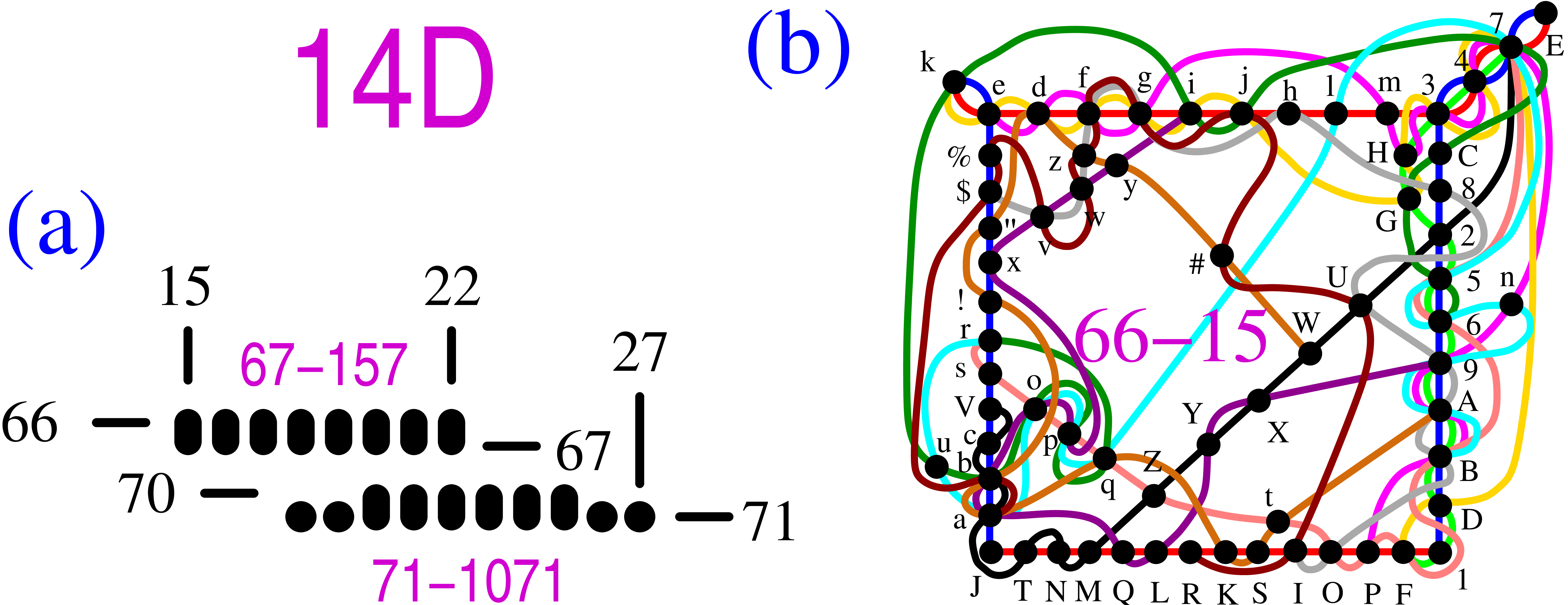}
\caption{(a) Distribution of 20,000 14D critical KS MMPHs
  obtained from the 67-157 and 71-1071 master MMPHs
  obtained via dimensional upscaling; Cf.~Fig.~\ref{fig:4d}(a);
  (b) the smallest critical KS MMPH 66-15 we obtained from the
  67-157 master; $HI_{cM}= \boldsymbol\alpha=6$, $HI_{cm}=2$;
  strings and coordinatizations of 66-15 are given in
  Appendix \ref{subsec:14ds}.}
\label{fig:14d}
\end{figure}

Spin-$\frac{13}{2}$ 14D criticals MMPHs shown in
Fig.~\ref{fig:14d}(a) are even fewer than the 13D ones,
again due to the generation complexity. Longer CPU time is not
viable at the present level of our research which is to show
that one can generate thousands of comparatively small
non-isomorphic criticals in high dimensional spaces. 
The smallest critical we obtained is 66-{\bf 15} in 
Fig.~\ref{fig:14d}(b).

$HI_{cm}=2$ for 66-15 means that it is, for instance, 
possible to assign 1 to just {\tt b} and {\tt B}. Then,
all other vertices must be assigned 0 and that violates
the condition (II) of the KS theorem for, e.g., the top
red hyperedge. Alternatively, a numerical verification can
be carried out on its MMPH string (Sec.~\ref{subsec:14ds})
proving that the (top red) hyperedge {\tt ekdfghijlm47E3}
contains neither {\tt b} nor {\tt B}.

\subsection{\label{subsec:15} 15D MMPHs---dimensional
                 upscaling}

\begin{figure}[h]
\center
\includegraphics[width=0.48\textwidth]{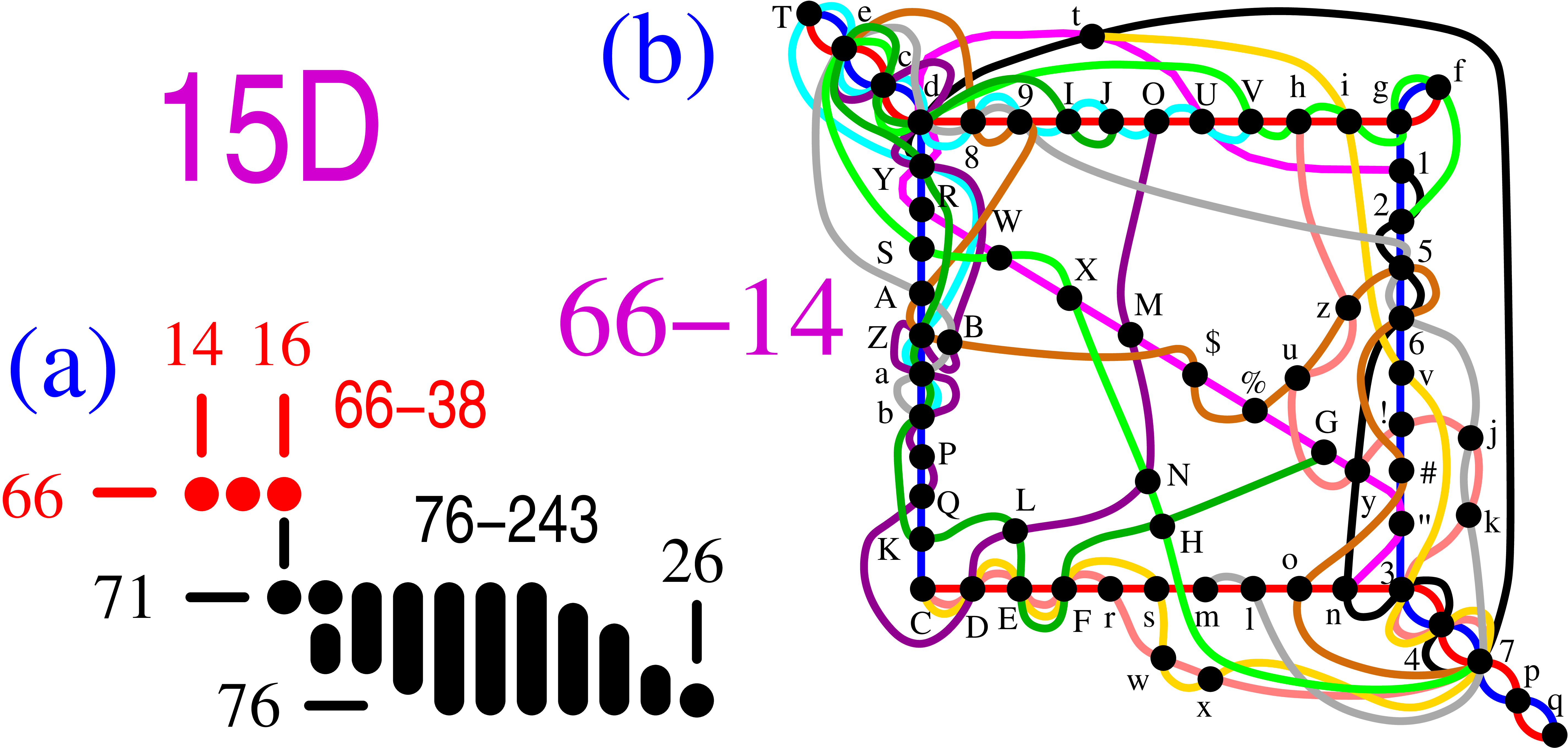}
\caption{(a) Distribution of 80,000 15D critical KS MMPHs
  obtained from the 66-38 and 76-243 master MMPHs;
  Cf.~Fig.~\ref{fig:4d}(a); (b) the smallest critical KS MMPH
  66-14 we obtained from the 66-38 master;
  $HI_{cM}=\boldsymbol\alpha=6$, $HI_{cm}=2$;
  strings and coordinatizations of 66-14
  are given in Appendix \ref{subsec:15ds}.}
\label{fig:15d}
\end{figure}

With spin-7 15D criticals MMPHs shown in Fig.~\ref{fig:15d} 
we succeeded in establishing an optimal dimensional upscaling.
So, the complexity of generation just allowed a more abundant
distribution than in the 14D space, while the minimal
MMPH has fewer hyperedges. The advantage of the low level
distribution is that we obtain fairly small criticals. Note
that the smallest obtained MMPH in the 15D space---66-{\bf 14}
shown in Fig.~\ref{fig:15d}(b)---has fewer hyperedges
than the ones in 13D and 14D obtained above; in the 16D they
are even fewer.

\subsection{\label{subsec:16} 16D MMPHs---dimensional
                 upscaling}

\begin{figure}[h]
\center
\includegraphics[width=0.48\textwidth]{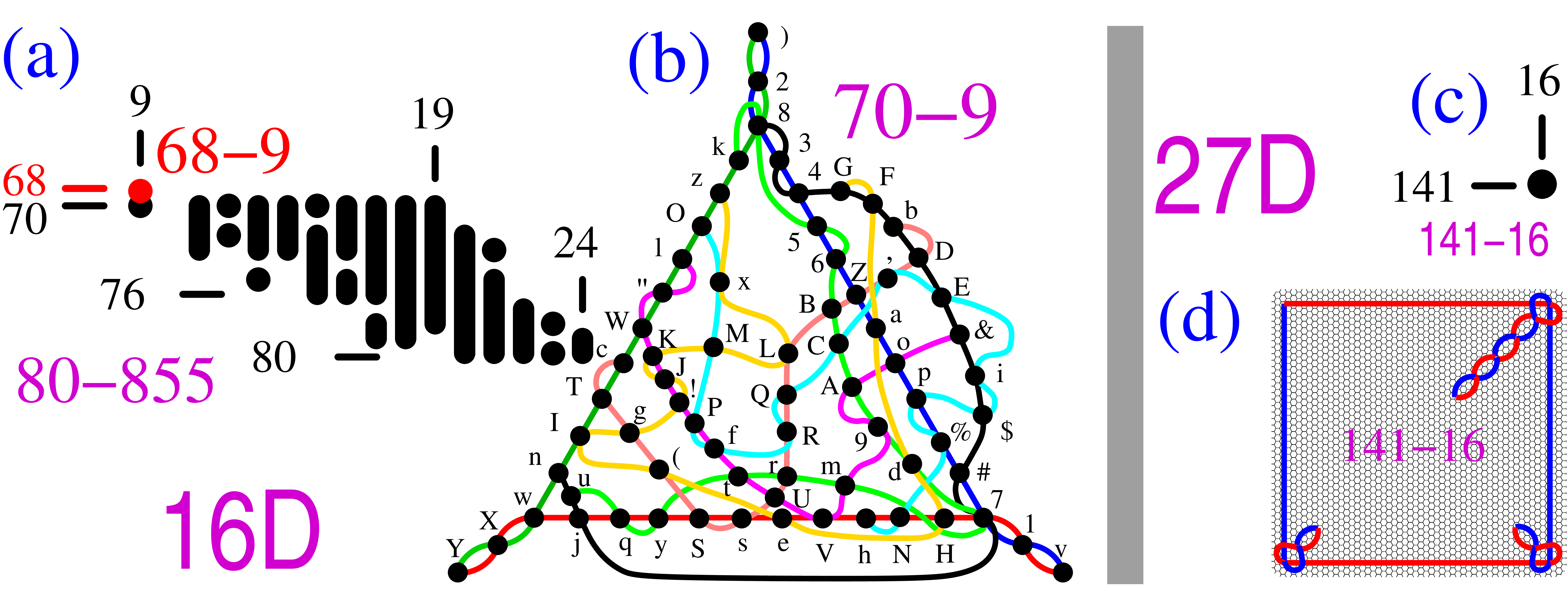}
\caption{(a) Distribution of 10,000 16D critical KS MMPHs obtained
  from the 80-855 master MMPH; Cf.~Fig.~\ref{fig:4d}(a); (b) the
  smallest critical KS MMPH 70-9 we obtained  from the master
  80-855; $HI_{cM}=\boldsymbol\alpha=4$, $HI_{cm}=3$; (c) critical
  141-16 is the only MMPH of the 141-16 master; (d) its biggest
  loop; $HI_{cM}= \boldsymbol\alpha=7$, $HI_{cm}=3$ strings and
  coordinatizations of 70-9 and 141-16 are given in
  Appendix \ref{subsec:16ds}.}
\label{fig:16d}
\end{figure}

16D space hosts four entangled qubits: $n=2^4$. The smallest
16D critical MMPH with nine hyperedges 70-{\bf 9} we obtained
via dimensional upscaling from the master 80-855, shown in
Fig.~\ref{fig:16d}(b), confirms our result from
\cite{waeg-aravind-pra-17} according to which $4n$D 
MMPHs (for positive integers $n$) require at most nine
hyperedges, and at most fifteen hyperedges in dimensions 4n+2. 

\subsection{\label{subsec:27} 27D MMPHs---dimensional
                 upscaling}

The 27D is the space of three entangled qutrits. We apply a
rank-scaling method to our 9D 47-16 to get a 27D 141-16
($3\times 47=141$) shown in Fig.~\ref{fig:16d}(c,d) (its
hyperedges are too interwoven to be discernible in (d);
therefore, only its biggest loop is given).
According to the hyperedge pattern that other MMPHs follow we
should be able to obtain a 27D MMPH with 14 hyperedges if we
only kept our computer search running long enough.

\subsection{\label{subsec:32} 32D MMPHs---dimensional
                 upscaling}

We are also able to straightforwardly generate 32D (five
entangled qubits) by combining two of the 16D 68-9s to get 32D
136-9 but did not do it here because the hypergraph is already
listed in \cite[Table V]{waeg-aravind-pra-17}, under $n=8N$,
34$N$-9 and because quite a number of 32D critical KS hypergraphs,
obtained by another method, is already given in
\cite[Sec.~XI]{pavicic-pra-17}.

\section{Applications}
\label{sec:app}

Exhaustive development of the applications presented here is
outside of the scope of this paper. However, we demonstrate
that any and all KS sets in the higher dimensions presented
below do have potential practical usage. 

\subsection{\label{appA}Larger alphabet}

We extend the lager alphabet procedure from \cite{bech-00}.
A 4D KS ``protection'' of quantum key distribution (QKD) protocols,
has been put forward in \cite{cabello-dambrosio-11}  based on a
modification of the BB84 protocol in \cite{svozil-10}.
A KS hypergraph with 9 edges has been used. (i) Alice randomly
picks one of 9 edges (bases) and sends Bob a randomly chosen state
(vertex) of that edge; (ii) Bob picks one of 9 edges at random 
and measures the system received from Alice. So, instead of 
qubits we deal with ququarts and the information transferred via a
system is not 1 but 2 bits \cite{bech-00}. We can modify and
generalize this QKD protocol so as to apply it to any $k$-$l$
hypergraph ($k$ vertices, $l$ edges). The probability of
Bob's picking a correct edge is $\frac{1}{l}$ and the number
of vertices rises linearly with dimension (see Table
\ref{T:small}). The higher the dimension, the more errors Eve
introduces. However, the main advantage of the KS QKD is that
the number of 1's one can assign to vertices is maximised
\cite[Def.~4.7, Lemma 4.8]{pavicic-quantum-23}
(Lov{\'a}sz's independence number $\boldsymbol\alpha$
\cite[p.~192]{gro-lovasz-schr-81}).
$\boldsymbol\alpha$'s for minimal hypergraphs 
in 9D,10D,\dots,16D,27D are given in Secs.~\ref{sec:disc}.

Ideally, if we obtained a higher number of 1's for any complete
set of data, Eve would be in the line. One can implement
the protocol by means of orbital angular momentums which reach over
$100\times 100$ entangled dimensionality \cite{krenn-zeilinger-14}.
Finally, the huge multiplicity of available hypergraphs enhances
the security of QKD, since an attacker may not know which protocol
is being used.

\subsection{\label{appB}Oblivious communication protocol
  and communication of bounded-dimensional systems
  protocols.}

The one-way
communication tasks presented in \cite{saha-hor-19}, which can make
use of any KS set, are communication of a system with bounded
dimension, and communication of a system with no dimensional bound,
but with some information about the sender's input unrevealed
(i.e., oblivious communication). The complete set of vertices
(quantum states) in the KS set (in the appropriate dimension) are
used as the input alphabet for both the sender's state preparation
and receiver's measurement setting. As a result, these quantum
communication protocols outperform the best corresponding classical
protocols for the same communication tasks. 

The protocol we have described is already completely
characterized in the research we have cited for lower dimensions,
and they demonstrate that any and all KS sets have practical
applications. Here, we show that the protocol can be
straightforwardly extended to any dimension given the 
necessary computational resources, which scale sub-exponentially
with the dimension.

\subsection{\label{appC}Generalized Hadamard matrices.}

Practically all known
quantum computation algorithms are based on Fourier transform of
which Hadamard ($H$) transform is a special case. Recently, a new
$S$ class of $H$ matrices ($S$-$H$) of even order in $\mathbb{C}^n$
has been designed for proving the existence of KS hypergraphs in
even $n$D. Our method generates any of these KS hypergraphs (which
all turn out to be star-like) and therefore---inversely---the
elements of the corresponding $S$-$H$ matrices.  

The new $S$ class of $H$ (Hadamard) $S$-$H$ matrices designed in 
\cite{lisonek-19} relies on the following theorem. 

Theorem (Lison{\v e}k 2019): Suppose that there exists an $S$-$H$
matrix of order $n$ ($n$ even); then there exists a KS
hypergraph $k$-$l$ in $\mathbb{C}^n$ such that $k=\binom{n+1}{2}$
and $l=n+1$

In \cite{lisonek-19} Lison{\v e}k defines a KS hypergraph in
$\mathbb{C}^n$ (Def.~1). By Def.~2.1 he defines an $S$-$H$ matrix
and in Def.~2.2 a generalized Hadamard matrix. Via
Theorem 3.1 he connects a KS hypergraph and a corresponding $S$-$H$
matrix. The proof provides us with an algorithm for a mutual mapping
of their elements in any even dimension. The details are outside
the scope of the present paper and we direct the reader to
Ref.~\cite{lisonek-19} for them. 

Only the 6D $21$-$7$ KS hypergraph was known to \cite{lisonek-19},
i.e., only one particular $S$-$H$ matrix (apart of the existence
of all of them). Our method generates any of these KS hypergraphs
together with their coordinatization. Inversely, and that is the
core of its application, it gives the elements of the corresponding
$S$-$H$ matrices. It also shows that all the corresponding KS
hypergraphs are star-like. The latter feature clarifies why $n$ has
to be even: one cannot draw a regular star with the Schl\"afli
symbol \{$n$+1/$\frac{n}{2}$\} in odd dimensional spaces because
$\frac{n}{2}$ has to be an integer. The 8D star 36-9 is given in
Sec.~\ref{subsec:8}. Note the $\{0,\pm1\}$ coordinatization.
The 10D star 55-11 is given in Sec.~\ref{subsec:10}.

\section{Discussion}
\label{sec:disc}

To summarize, we design a feasible unified
method of generating quantum contextual sets in higher dimensions
as a response to recent calls for high dimensional contextual
applications in quantum computation and quantum communication.
The need for such a unified method appeared because previous
particular attempts were scattered, and the only previously existing
unifying method was of exponential complexity and therefore not
applicable to 9+ dimensions. This is all presented in detail
with references in the Introduction. 

Because the complexity of small KS sets does not scale with
dimension, the computational resources needed for our method
do not generally scale exponentially, depending on the type
of new KS sets one is looking for. It is based on dimensional
upscaling, i.e., on obtaining higher dimensional contextual
sets from previously obtained ones in lower dimensions.
Its goal is to generate manageable-sized master KS hypergraph
wherefrom we can obtain a large number of non-isomorphic KS
hypergraphs in high dimensional spaces for any possible future
application and implementation, e.g., in quantum computation and 
communication, parity oblivious transfer, genuine random number
generation, quantum dimension certification, and relational
database theory, three of which are presented in
Sec.~\ref{sec:app}. A list of the smallest obtained KS
hypergraphs is given in Table~\ref{T:small}. 

Our preliminary testings show that KS hypergraphs in dimensions
much higher than 32 can be generated depending on the amount of
the CPU time one is ready to sacrifice on supercomputers. 

\begin{table}[ht]
\caption{The smallest KS hypergraphs obtained by our methods.
  A pattern appears to emerge wherein the minimum number of
  edges per hypergraph of dimension $4n+(0,1,2,3)$ are
  $(9,16,15,14)$ for integers $n\geq 1$, which confirms that
  the minimum complexity of KS hypergraphs does not grow with
  dimension;  distributions are given in Sec.~\ref{sec:review}
  and MMPH strings and coordinatizations are given in
  Appendix.}\center
\setlength{\tabcolsep}{3.6pt}
\begin{tabular}{|c|ccccc|}
    \hline
  \multirow{3}{*}{dim\ } & Smallest &
  \multirow{3}{*}{$\boldsymbol\alpha$\hskip-7pt\rotatebox{90}{\ \ \ Lov{\'a}sz\hskip-2pt}}&  
  No.~of 
  & \multirow{2}*{Smallest}
  & \multirow{2}*{Vector} \\
&  critical &  &  non-isom & \multirow{2}*{master} &
      \multirow{2}*{components} \\
&hypergraphs   &     & smallest    &  &    \\
  \hline
  {4D}&\hbox to 30pt{}18-{\bf 9}& 4 & 1 
   &24-24& $\{0,\pm 1\}$\\
  {5D}&29-{\bf 16}& 7 & 2 
   &105-136& $\{0,\pm 1\}$\\
  {6D}&\hbox to 10pt{}31-{\bf 15}& 7& 1 
   &31-16& $\{0,\pm 1\}$\\
  {7D}&\hbox to 20pt{}34-{\bf 14}& 7& 2 
   &47-176& $\{0,\pm 1\}$\\
  {8D}&\hbox to 30pt{}34-{\bf 9}& 4& 2 
   &120-2024& $\{0,\pm 1\}$\\
  {9D} &47-{\bf 16}& 7& 2  & 47-144 & $\{0,\pm 1\}$ \\
   {10D}&\hbox to 10pt{}50-{\bf 15}& 7 &66  & 52-141 &
   $\{0,\pm 1\}$ \\
   {11D}&\hbox to 20pt{}50-{\bf 14}& 6 &1603 & 50-38 &
   $\{0,\pm 1\}$ \\
   {12D}&\hbox to 30pt{}52-{\bf 9}& 4 &502 & 52-81 &
   $\{0,\pm 1\}$ \\
   {13D}&63-{\bf 16}& 7 &23 & 63-144 &
   $\{0,\pm 1\}$ \\
   {14D}&\hbox to 10pt{}66-{\bf 15}& 6 &17 & 67-157 &  
   $\{0,\pm 1\}$ \\
   {15D}&\hbox to 20pt{}66-{\bf 14}& 6 &177 & 66-38 &
   $\{0,\pm 1\}$ \\
   {16D}&\hbox to 30pt{}68-{\bf 9}& 4 &1 & 68-9 & 
   $\{0,\pm 1\}$ \\
   {\dots}&\dots&\dots & \dots & \dots & 
   \dots \\
   {27D}&\hbox to 13pt{}141-16& 7 &1 & 141-16 &
   $\{0,\pm 1\}$ \\
  \hline\end{tabular}
\label{T:small}
\end{table}

\medskip
\smallskip
{\em Acknowledgements}---Supported by the Ministry of Science,
Education, and Youth of Croatia through Center of Excellence CEMS
funding, and by the European structural and investment funds (ESIF),
MSE Grant No. KK.01.1.1.01.0001.
Computational support was provided by the Zagreb University
Computing Centre. Programs are freely available from our
repository \cite{puh-repository}.

\vfill

\appendix

\renewcommand{\thesection}{}
\renewcommand{\thesubsection}{\arabic{subsection}}

\begin{widetext}

\section{KS MMPH's strings and coordinatizations}

Below we provide strings and coordinatizations of KS MMPH's
referred to in Sections IV.B-O.

\subsection{\label{subsec:4ds}4D MMPHs; 9 hyperedges (18-9)}

{\bf 18-9} {\tt 1234,4567,789A,ABCD,DEFG,GHI1,35EC,29BI,68FH.} {\tt 1}=(0,0,0,1), {\tt 2}=(0,0,1,0), {\tt 3}=(1,1,0,0), {\tt 4}=(1,-1,0,0), {\tt 5}=(0,0,1,1), {\tt 6}=(1,1,1,-1), {\tt 7}=(1,1,-1,1), {\tt 8}=(1,-1,1,1), {\tt 9}=(1,0,0,-1), {\tt A}=(0,1,1,0), {\tt B}=(1,0,0,1), {\tt C}=(1,-1,1,-1),\break {\tt D}=(1,1,-1,-1),\quad {\tt E}=(1,-1,-1,1),\quad  {\tt F}=(0,1,0,1),\quad  {\tt G}=(1,0,1,0),\quad  {\tt H}=(1,0,-1,0),\quad  {\tt I}=(0,1,0,0)

\medskip

{\bf 24-24} {\tt LMNO,HIJK,DEFG,BCFG,9ADE,78EG,56DF,5678,9ABC,68JK,57HI,ACIK,9BHJ,1234,4DGO,3EFN,258M,167L,\break 19CM,2ABL,3HKO,4IJN,34NO,12LM.} {\tt 1}=(0,0,0,1), {\tt 2}=(0,0,1,0), {\tt 3}=(1,1,0,0), {\tt 4}=(1,-1,0,0), {\tt 5}=(0,1,0,-1), {\tt 6}=(1,0,-1,0), {\tt 7}=(1,0,1,0), {\tt 8}=(0,1,0,1), {\tt 9}=(0,1,-1,0), {\tt A}=(1,0,0,-1), {\tt B}=(1,0,0,1), {\tt C}=(0,1,1,0), {\tt D}=(1,1,1,1), {\tt E}=(1,-1,-1,1), {\tt F}=(1,-1,1,-1), {\tt G}=(1,1,-1,-1), {\tt H}=(-1,1,1,1), {\tt I}=(1,1,-1,1), {\tt J}=(1,1,1,-1), {\tt K}=(1,-1,1,1), {\tt L}=(0,1,0,0), {\tt M}=(1,0,0,0), {\tt N}=(0,0,1,1), {\tt O}=(0,0,1,-1)

\subsection{\label{subsec:5ds}5D MMPH;  16 hyperedges}

{\bf 29-16} {\tt HOINJ,JGSTF,FT4Q8,85679,92LOH,PQRST,KLMNO,CDEIO,ABEGT,34DMO,12BRT,237CO,146AT,279OP,468KT,\break 5EJOT.} {\tt 1}=(1,-1,1,0,-1), {\tt 2}=(1,0,-1,0,0), {\tt 3}=(1,-1,1,1,0), {\tt 4}=(0,1,1,0,0), {\tt 5}=(0,0,1,0,0), {\tt 6}=(1,0,0,0,1), {\tt 7}=(0,1,0,1,0), {\tt 8}=(1,0,0,0,-1), {\tt 9}=(0,1,0,-1,0), {\tt A}=(1,1,-1,0,-1), {\tt B}=(1,1,1,0,1), {\tt C}=(1,1,1,-1,0), {\tt D}=(1,1,-1,1,0), {\tt E}=(1,-1,0,0,0),\break {\tt F}=(1,-1,1,0,1), {\tt G}=(0,0,1,0,-1), {\tt H}=(1,-1,1,-1,0), {\tt I}=(0,0,1,1,0), {\tt J}=(1,1,0,0,0), {\tt K}=(0,1,-1,0,0), {\tt L}=(1,1,1,1,0),\break {\tt M}=(1,0,0,-1,0), {\tt N}=(1,-1,-1,1,0), {\tt O}=(0,0,0,0,1), {\tt P}=(1,0,1,0,0), {\tt Q}=(1,1,-1,0,1), {\tt R}=(0,1,0,0,-1), {\tt S}=(-1,1,1,0,1), {\tt T}=(0,0,0,1,0)

\subsection{\label{subsec:6ds}6D MMPHs; 15 to 17 hyperedges}
  
{\bf 31-15} {\tt A78B91,194CD3,3GK65J,J5HLME,EFQSTA,123456,2EF9AB,GHI856,NOI756,NP4K56,OPFQR6,OPFLM5,EFRSCU,\break EFVLDU,EFVMBT.} {\tt 1}=(0,1,0,0,0,0); {\tt 2}=(0,0,1,0,0,0); {\tt G}=(1,1,1,1,0,0); {\tt H}=(1,-1,1,-1,0,0); {\tt J}=(1,-1,-1,1,0,0);\break  {\tt N}=(1,-1,-1,-1,0,0); {\tt O}=(1,1,-1,1,0,0); {\tt P}=(1,1,1,-1,0,0); {\tt E}=(1,1,0,0,0,0); {\tt F}=(1,-1,0,0,0,0); {\tt I}=(0,1,0,-1,0,0); {\tt 7}=(1,0,1,0,0,0); {\tt 8}=(1,0,-1,0,0,0); {\tt 3}=(1,0,0,-1,0,0); {\tt 4}=(1,0,0,1,0,0); {\tt K}=(0,1,-1,0,0,0); {\tt 9}=(0,0,0,0,1,0); {\tt Q}=(0,0,1,1,1,1); {\tt R}=(0,0,1,1,-1,-1); {\tt S}=(0,0,1,-1,1,-1); {\tt V}=(0,0,1,-1,1,1); {\tt L}=(0,0,1,1,-1,1); {\tt M}=(0,0,1,1,1,-1); {\tt 5}=(0,0,0,0,1,1);\break  {\tt 6}=(0,0,0,0,1,-1); {\tt C}=(0,0,1,0,0,1); {\tt D}=(0,0,1,0,0,-1); {\tt U}=(0,0,0,1,1,0); {\tt A}=(0,0,0,1,0,-1); {\tt B}=(0,0,0,1,0,1);\break  {\tt T}=(0,0,1,0,-1,0).

\medskip
{\bf 32-16} {\tt 123456,123789,12ABCD,12BE9F,12ECG5,12HIG6,12IJ8F,13KL47,23MN47,OPLQ47,ORMS47,PRTBCD,UVTA47,\break UVTHJD,UWNS47,VWKQ47.} {\tt 1}=(1,0,0,0,0,0); {\tt 2}=(0,1,0,0,0,0); {\tt 3}=(0,0,1,0,0,0); {\tt O}=(1,1,1,1,0,0);\break  {\tt P}=(1,-1,1,-1,0,0); {\tt R}=(1,-1,-1,1,0,0); {\tt U}=(1,-1,-1,-1,0,0); {\tt V}=(1,-1,1,1,0,0); {\tt W}=(1,1,1,-1,0,0); {\tt T}=(1,1,0,0,0,0); {\tt A}=(0,0,1,-1,0,0); {\tt K}=(0,1,0,1,0,0); {\tt L}=(0,1,0,-1,0,0); {\tt Q}=(1,0,-1,0,0,0); {\tt M}=(1,0,0,-1,0,0); {\tt N}=(1,0,0,1,0,0);\break  {\tt S}=(0,1,-1,0,0,0); {\tt 4}=(0,0,0,0,1,0); {\tt 7}=(0,0,0,0,0,1); {\tt B}=(0,0,1,1,1,1); {\tt E}=(0,0,1,-1,-1,1); {\tt C}=(0,0,1,1,-1,-1);\break  {\tt H}=(0,0,1,-1,-1,-1); {\tt I}=(0,0,1,1,-1,1); {\tt J}=(0,0,1,-1,1,1); {\tt G}=(0,0,1,0,1,0); {\tt 5}=(0,0,0,1,0,1); {\tt 6}=(0,0,0,1,0,-1); {\tt 8}=(0,0,0,1,1,0); {\tt 9}=(0,0,0,1,-1,0); {\tt F}=(0,0,1,0,0,-1); {\tt D}=(0,0,0,0,1,-1).

\medskip
{\bf 33-17} (one of four) {\tt 123456,123789,12ABCD,12EFGD,12FH9I,12HGJ5,12BKJ6,12KC8I,13LM47,23NO47,PQMR47,\break PSNT47,QSUA47,QSUFGD,VWUE47,VXOT47,WXLR47.} {\tt 1}=(1,0,0,0,0,0); {\tt 2}=(0,1,0,0,0,0); {\tt 3}=(0,0,1,0,0,0); {\tt P}=(1,1,1,1,0,0); {\tt Q}=(1,-1,1,-1,0,0); {\tt S}=(1,-1,-1,1,0,0); {\tt V}=(1,-1,-1,-1,0,0); {\tt W}=(1,-1,1,1,0,0); {\tt X}=(1,1,1,-1,0,0); {\tt U}=(1,1,0,0,0,0); {\tt A}=(0,0,1,1,0,0); {\tt E}=(0,0,1,-1,0,0); {\tt L}=(0,1,0,1,0,0); {\tt M}=(0,1,0,-1,0,0); {\tt R}=(1,0,-1,0,0,0); {\tt N}=(1,0,0,-1,0,0); {\tt O}=(1,0,0,1,0,0); {\tt T}=(0,1,-1,0,0,0); {\tt 4}=(0,0,0,0,1,0); {\tt 7}=(0,0,0,0,0,1); {\tt F}=(0,0,1,1,1,1); {\tt H}=(0,0,1,-1,-1,1); {\tt G}=(0,0,1,1,-1,-1);\break  {\tt B}=(0,0,1,-1,-1,-1); {\tt K}=(0,0,1,1,-1,1); {\tt C}=(0,0,1,-1,1,1); {\tt J}=(0,0,1,0,1,0); {\tt 5}=(0,0,0,1,0,1); {\tt 6}=(0,0,0,1,0,-1); {\tt 8}=(0,0,0,1,1,0); {\tt 9}=(0,0,0,1,-1,0); {\tt I}=(0,0,1,0,0,-1); {\tt D}=(0,0,0,0,1,-1).

\subsection{\label{subsec:7ds}7D MMPH; 14 hyperedges}

{\bf 34-14} {\tt 1234567, 189A5BC, 189DE7F, 189GHIJ, 189KHBL, 2MNDOIP, 2MNEOCL, 2MNGK6F, QRNSAJP, QT4U567, RTV9567, WXMS567, WYV8AJP, XY3U567.} {\tt 1}=(0,0,0,1,0,0,0); {\tt 2}=(0,0,1,0,0,0,0); {\tt 3}=(1,-1,0,0,0,0,0); {\tt 4}=(1,1,0,0,0,0,0); {\tt 5}=(0,0,0,0,0,0,1); {\tt 6}=(0,0,0,0,1,1,0); {\tt 7}=(0,0,0,0,1,-1,0); {\tt 8}=(0,1,-1,0,0,0,0); {\tt 9}=(0,1,1,0,0,0,0); {\tt A}=(0,0,0,0,1,0,0); {\tt B}=(1,0,0,0,0,-1,0); {\tt C}=(1,0,0,0,0,1,0); {\tt D}=(1,0,0,0,1,1,-1); {\tt E}=(-1,0,0,0,1,1,1); {\tt F}=(1,0,0,0,0,0,1); {\tt G}=(1,0,0,0,1,-1,-1); {\tt H}=(1,0,0,0,1,1,1); {\tt I}=(1,0,0,0,-1,0,0); {\tt J}=(0,0,0,0,0,1,-1); {\tt K}=(1,0,0,0,-1,1,-1); {\tt L}=(0,0,0,0,1,0,-1); {\tt M}=(0,1,0,1,0,0,0); {\tt N}=(0,1,0,-1,0,0,0); {\tt O}=(1,0,0,0,1,-1,1); {\tt P}=(0,0,0,0,0,1,1); {\tt Q}=(-1,1,1,1,0,0,0); {\tt R}=(1,1,-1,1,0,0,0); {\tt S}=(1,0,1,0,0,0,0); {\tt T}=(1,-1,1,1,0,0,0); {\tt U}=(0,0,1,-1,0,0,0); {\tt V}=(1,0,0,-1,0,0,0); {\tt W}=(1,-1,-1,1,0,0,0); {\tt X}=(1,1,-1,-1,0,0,0); {\tt Y}=(1,1,1,1,0,0,0).

\subsection{\label{subsec:8ds}8D MMPH; 9 hyperedges}

{\bf 34-9} {\tt 12345678,9ABCDEFG,HIJKLMFG,NOPQME78,RSTULD56,VWXUKC46,XTPQJB36,YVWOIA28,YRSNH918.} {\tt 1}=(0,0,1,0,0,0,0,0), {\tt 2}=(1,0,0,0,0,0,0,0), {\tt 3}=(0,0,0,0,0,1,0,0), {\tt 4}=(0,0,0,0,1,0,0,0), {\tt 5}=(0,0,0,1,0,0,0,0), {\tt 6}=(0,0,0,0,0,0,0,1), {\tt 7}=(0,1,0,0,0,0,0,0), {\tt 8}=(0,0,0,0,0,0,1,0), {\tt 9}=(1,1,0,-1,0,0,0,-1), {\tt A}=(0,1,-1,0,-1,0,0,1),\break  {\tt B}=(0,1,0,1,1,0,-1,0), {\tt C}=(1,0,0,1,0,-1,1,0), {\tt D}=(0,0,-1,0,1,1,1,0), {\tt E}=(1,0,1,0,0,1,0,1), {\tt F}=(-1,1,1,0,0,0,1,0),\break  {\tt G}=(0,0,0,1,-1,1,0,-1), {\tt H}=(1,1,0,1,0,0,0,1), {\tt I}=(0,1,-1,0,1,0,0,-1), {\tt J}=(0,-1,0,1,1,0,1,0), {\tt K}=(1,0,0,-1,0,1,1,0), {\tt L}=(0,0,1,0,1,1,-1,0), {\tt M}=(1,0,1,0,0,-1,0,-1), {\tt N}=(0,0,0,1,1,1,0,-1), {\tt O}=(0,0,0,1,1,-1,0,1), {\tt P}=(0,0,0,1,-1,0,0,0),\break  {\tt Q}=(1,0,-1,0,0,0,0,0), {\tt R}=(0,0,0,0,1,-1,0,0), {\tt S}=(1,-1,0,0,0,0,0,0), {\tt T}=(1,1,1,0,0,0,1,0), {\tt U}=(1,1,-1,0,0,0,-1,0), {\tt V}=(0,0,0,1,0,1,0,0),\qquad {\tt W}=(0,1,1,0,0,0,0,0),\qquad  {\tt X}=(1,-1,1,0,0,0,-1,0),\qquad  {\tt Y}=(0,0,0,-1,1,1,0,1)

\medskip
{\bf 36-9} {\tt 12345678,89ABCDEF,FGHI4JKL,L7MNBOPQ,QERSI3TU,UK6VNAWX,XPDYSH2Z,ZTJ5VM9a,aWOCYRG1.} {\tt 1}=(0,0,0,0,0,0,0,1), {\tt 2}=(0,0,0,0,0,0,1,0), {\tt 3}=(0,0,0,0,0,1,0,0), {\tt 4}=(0,0,0,0,1,0,0,0), {\tt 5}=(0,0,1,1,0,0,0,0),\break {\tt 6}=(0,0,1,-1,0,0,0,0), {\tt 7}=(1,1,0,0,0,0,0,0), {\tt 8}=(1,-1,0,0,0,0,0,0), {\tt 9}=(1,1,0,0,0,0,-1,1), {\tt A}=(0,0,1,1,1,-1,0,0), {\tt B}=(0,0,0,0,0,0,1,1), {\tt C}=(0,0,1,-1,1,1,0,0), {\tt D}=(0,0,0,1,0,1,0,0), {\tt E}=(0,0,1,0,-1,0,0,0), {\tt F}=(1,1,0,0,0,0,1,-1),\break {\tt G}=(0,0,1,0,0,-1,0,0), {\tt H}=(1,0,0,0,0,0,0,1), {\tt I}=(0,0,0,1,0,0,0,0), {\tt J}=(1,-1,0,0,0,0,-1,-1), {\tt K}=(0,1,0,0,0,0,-1,0), {\tt L}=(0,0,1,0,0,1,0,0), {\tt M}=(0,0,1,-1,1,-1,0,0), {\tt N}=(1,-1,0,0,0,0,-1,1), {\tt O}=(0,0,0,1,1,0,0,0), {\tt P}=(0,0,1,1,-1,-1,0,0),\break {\tt Q}=(1,-1,0,0,0,0,1,-1), {\tt R}=(1,0,0,0,0,0,-1,0), {\tt S}=(0,0,1,0,1,0,0,0), {\tt T}=(0,1,0,0,0,0,0,-1), {\tt U}=(1,1,0,0,0,0,1,1), {\tt V}=(0,0,0,0,1,1,0,0), {\tt W}=(0,0,1,1,-1,1,0,0), {\tt X}=(1,0,0,0,0,0,0,-1), {\tt Y}=(0,1,0,0,0,0,0,0), {\tt Z}=(0,0,1,-1,-1,1,0,0), {\tt a}=(1,0,0,0,0,0,1,0)

\subsection{\label{subsec:9ds}9D MMPH; 16 hyperedges}

{\bf 47-16} {\tt 234567891,1BOPERSNQ,QdgXJDaFY,YaFZIbc32,12ABC5DEF,13GHIJ7KF,1A4567LMN,1G4567STU,1HOV6KWLU, 1XOV6KRT9,1XIJ7KWM8,deHfOV6KF,eghA4567F,ijhBOPEQF,ikZXPVblF,jkGfC4lcF.}
{\tt 1}=(1,0,0,0,0,0,0,0,0); {\tt 2}=(0,1,0,0,0,0,0,0,0); {\tt 3}=(0,0,1,0,0,0,0,0,0); {\tt d}=(1,1,1,1,0,0,0,0,0); {\tt e}=(1,-1,1,-1,0,0,0,0,0); {\tt g}=(1,-1,-1,1,0,0,0,0,0);\break {\tt i}=(1,-1,-1,-1,0,0,0,0,0); {\tt j}=(1,-1,1,1,0,0,0,0,0); {\tt k}=(1,1,1,-1,0,0,0,0,0); {\tt h}=(1,1,0,0,0,0,0,0,0); {\tt A}=(0,0,1,1,0,0,0,0,0); {\tt B}=(0,0,1,-1,0,0,0,0,0); {\tt G}=(0,1,0,1,0,0,0,0,0); {\tt H}=(0,1,0,-1,0,0,0,0,0); {\tt f}=(1,0,-1,0,0,0,0,0,0); {\tt Y}=(1,0,0,-1,0,0,0,0,0); {\tt Z}=(1,0,0,1,0,0,0,0,0); {\tt X}=(0,1,-1,0,0,0,0,0,0); {\tt C}=(0,0,0,0,1,0,0,0,0); {\tt 4}=(0,0,0,0,0,1,0,0,0); {\tt 5}=(0,0,0,0,0,0,1,0,0); {\tt O}=(0,0,0,0,1,1,1,1,0); {\tt P}=(0,0,0,0,1,-1,1,-1,0); {\tt V}=(0,0,0,0,1,-1,-1,1,0); {\tt I}=(0,0,0,0,1,-1,-1,-1,0); {\tt a}=(0,0,0,0,1,-1,1,1,0); {\tt J}=(0,0,0,0,1,1,1,-1,0); {\tt b}=(0,0,0,0,1,1,0,0,0); {\tt l}=(0,0,0,0,0,0,1,1,0); {\tt c}=(0,0,0,0,0,0,1,-1,0); {\tt D}=(0,0,0,0,0,1,0,1,0); {\tt E}=(0,0,0,0,0,1,0,-1,0); {\tt Q}=(0,0,0,0,1,0,-1,0,0); {\tt 6}=(0,0,0,0,1,0,0,-1,0); {\tt 7}=(0,0,0,0,1,0,0,1,0); {\tt K}=(0,0,0,0,0,1,-1,0,0); {\tt F}=(0,0,0,0,0,0,0,0,1); {\tt W}=(0,1,1,1,0,0,0,0,1); {\tt L}=(0,1,-1,1,0,0,0,0,-1); {\tt M}=(0,1,1,-1,0,0,0,0,-1); {\tt R}=(0,1,1,1,0,0,0,0,-1); {\tt S}=(0,1,-1,-1,0,0,0,0,-1); {\tt T}=(0,1,1,-1,0,0,0,0,1); {\tt N}=(0,1,0,0,0,0,0,0,1); {\tt U}=(0,0,1,0,0,0,0,0,-1); {\tt 8}=(0,0,0,1,0,0,0,0,-1); {\tt 9}=(0,0,0,1,0,0,0,0,1)

\subsection{\label{subsec:10ds}10D MMPH; 11 and 15 hyperedges}

{Star-like$\,\,\,$\bf 55-11} {\tt 123456789A,ABCDEFGHIJ,JKLMN5OPQR,R9STUEVWXY,YIZabN4cde,eQ8fgUDhij,jXHklbM3mn,\break ndP7ogTCpq,qiWGrlaL2s,smcO6ofSBt,tphVFrkZK1.}

\medskip
{\bf 50-15} {\tt 2ajkBCJST1,1TEOPRbmlD,Dl89FghnVU,UV46IceoqG,GHLNXZajk2,1DEJXYdeij,1DELMVWajk,2GHJXYfhpq, 2GHMNRSblm,2GHOQYZblm,2GHPQUWajk,45EFUVabop,56ABUVcdmn,78ACUVfgik,79HIJSTblm.} {\tt 1}=(1,0,0,0,0,0,0,0,0,0); {\tt 2}=(0,1,0,0,0,0,0,0,0,0); {\tt 4}=(1,1,1,1,0,0,0,0,0,0); {\tt 5}=(1,-1,1,-1,0,0,0,0,0,0); {\tt 6}=(1,-1,-1,1,0,0,0,0,0,0);\break {\tt 7}=(1,-1,-1,-1,0,0,0,0,0,0); {\tt 8}=(1,-1,1,1,0,0,0,0,0,0); {\tt 9}=(1,1,1,-1,0,0,0,0,0,0); {\tt A}=(1,1,0,0,0,0,0,0,0,0); {\tt B}=(0,0,1,1,0,0,0,0,0,0); {\tt C}=(0,0,1,-1,0,0,0,0,0,0); {\tt D}=(0,1,0,1,0,0,0,0,0,0); {\tt E}=(0,1,0,-1,0,0,0,0,0,0);\break {\tt F}=(1,0,-1,0,0,0,0,0,0,0); {\tt G}=(1,0,0,-1,0,0,0,0,0,0); {\tt H}=(1,0,0,1,0,0,0,0,0,0); {\tt I}=(0,1,-1,0,0,0,0,0,0,0); {\tt J}=(0,0,0,0,1,0,0,0,0,0); {\tt L}=(0,0,1,0,1,1,1,0,0,0); {\tt M}=(0,0,1,0,-1,1,-1,0,0,0); {\tt N}=(0,0,1,0,-1,-1,1,0,0,0);\break {\tt O}=(0,0,1,0,-1,-1,-1,0,0,0); {\tt P}=(0,0,1,0,-1,1,1,0,0,0); {\tt Q}=(0,0,1,0,1,1,-1,0,0,0); {\tt R}=(0,0,1,0,1,0,0,0,0,0); {\tt S}=(0,0,0,0,0,1,1,0,0,0); {\tt T}=(0,0,0,0,0,1,-1,0,0,0); {\tt U}=(0,0,0,0,1,0,1,0,0,0); {\tt V}=(0,0,0,0,1,0,-1,0,0,0);\break {\tt W}=(0,0,1,0,0,-1,0,0,0,0); {\tt X}=(0,0,1,0,0,0,-1,0,0,0); {\tt Y}=(0,0,1,0,0,0,1,0,0,0); {\tt Z}=(0,0,0,0,1,-1,0,0,0,0);\break {\tt a}=(0,0,0,0,0,0,0,1,0,0); {\tt b}=(0,0,0,0,0,0,0,0,1,0); {\tt c}=(0,0,0,0,0,1,0,1,1,1); {\tt d}=(0,0,0,0,0,1,0,-1,1,-1);\break {\tt e}=(0,0,0,0,0,1,0,-1,-1,1); {\tt f}=(0,0,0,0,0,1,0,-1,-1,-1); {\tt g}=(0,0,0,0,0,1,0,-1,1,1); {\tt h}=(0,0,0,0,0,1,0,1,1,-1);\break {\tt i}=(0,0,0,0,0,1,0,1,0,0); {\tt j}=(0,0,0,0,0,0,0,0,1,1); {\tt k}=(0,0,0,0,0,0,0,0,1,-1); {\tt l}=(0,0,0,0,0,0,0,1,0,1);\break {\tt m}=(0,0,0,0,0,0,0,1,0,-1); {\tt n}=(0,0,0,0,0,1,0,0,-1,0); {\tt o}=(0,0,0,0,0,1,0,0,0,-1); {\tt p}=(0,0,0,0,0,1,0,0,0,1);\break {\tt q}=(0,0,0,0,0,0,0,1,-1,0)

\subsection{\label{subsec:11ds}11D MMPH; 14 hyperedges}

{\bf 50-14} {\tt 567E234CFD1,1DGHJKLMANI,IYlmnUk9QXe,ecdfgOW67E5,123456789AB,1GHIJKLOPQR,27STUVKL8FW,\break 27STUVKL9QX,347YZabMDAN,567cdefMCXR,567cdefgPBN,cdhijkJV8FW,fZoHTjmn8FW,abGShilo8FW.} {\tt 1}=(0,0,1,1,1,1,0,0,0,0,0); {\tt 2}=(0,0,1,-1,1,-1,0,0,0,0,0); {\tt 3}=(0,0,0,1,0,-1,0,0,0,0,0); {\tt 4}=(0,0,1,0,-1,0,0,0,0,0,0); {\tt 5}=(0,1,0,0,0,0,0,0,0,0,0); {\tt 6}=(1,0,0,0,0,0,0,0,0,0,0); {\tt 7}=(0,0,0,0,0,0,1,0,0,0,0); {\tt 8}=(0,0,0,1,0,0,0,0,0,0,0); {\tt 9}=(0,0,1,0,0,0,0,0,0,0,0); {\tt A}=(0,0,0,0,0,1,0,0,0,0,0); {\tt B}=(0,0,0,0,1,0,0,0,0,0,0); {\tt C}=(1,-1,1,0,1,0,0,0,0,0,0); {\tt D}=(1,1,0,1,0,1,0,0,0,0,0); {\tt E}=(1,1,0,-1,0,-1,0,0,0,0,0); {\tt F}=(-1,1,1,0,1,0,0,0,0,0,0); {\tt G}=(0,1,-1,1,0,0,1,0,0,0,0); {\tt H}=(1,0,1,1,0,0,0,-1,0,0,0); {\tt I}=(1,0,0,0,1,1,0,1,0,0,0); {\tt J}=(0,1,0,0,-1,1,-1,0,0,0,0); {\tt K}=(0,0,1,0,-1,0,1,1,0,0,0); {\tt L}=(0,0,0,1,0,-1,-1,1,0,0,0); {\tt M}=(1,0,1,0,0,-1,1,0,0,0,0); {\tt N}=(0,-1,1,0,0,1,0,1,0,0,0); {\tt O}=(-1,1,0,0,0,0,1,1,0,0,0); {\tt P}=(1,0,-1,-1,0,0,0,1,0,0,0); {\tt Q}=(0,1,1,-1,0,0,-1,0,0,0,0); {\tt R}=(1,0,0,1,-1,0,-1,0,0,0,0); {\tt S}=(0,1,0,1,1,0,0,1,0,0,0); {\tt T}=(1,1,0,0,0,0,1,-1,0,0,0); {\tt U}=(0,1,0,0,1,-1,-1,0,0,0,0); {\tt V}=(1,0,0,0,-1,-1,0,1,0,0,0); {\tt W}=(1,1,0,-1,0,1,0,0,0,0,0);\break {\tt X}=(1,-1,-1,0,1,0,0,0,0,0,0); {\tt Y}=(0,0,0,0,0,0,0,0,1,0,0); {\tt Z}=(0,0,0,0,0,0,0,0,0,1,0); {\tt a}=(0,0,0,0,0,0,0,1,1,1,1); {\tt b}=(0,0,0,0,0,0,0,1,-1,1,-1); {\tt c}=(0,0,0,0,0,0,0,1,-1,-1,1); {\tt d}=(0,0,0,0,0,0,0,1,-1,-1,-1); {\tt e}=(0,0,0,0,0,0,0,1,-1,1,1); {\tt f}=(0,0,0,0,0,0,0,1,1,1,-1); {\tt g}=(0,0,0,0,0,0,0,1,1,0,0); {\tt h}=(0,0,0,0,0,0,0,0,0,1,1); {\tt i}=(0,0,0,0,0,0,0,0,0,1,-1); {\tt j}=(0,0,0,0,0,0,0,0,1,0,1); {\tt k}=(0,0,0,0,0,0,0,0,1,0,-1); {\tt l}=(0,0,0,0,0,0,0,1,0,-1,0); {\tt m}=(0,0,0,0,0,0,0,1,0,0,-1); {\tt n}=(0,0,0,0,0,0,0,1,0,0,1);\qquad {\tt o}=(0,0,0,0,0,0,0,0,1,-1,0)

\subsection{\label{subsec:12ds}12D MMPH; 9 hyperedges}
\vskip-10pt
{\bf 52-9} {\tt UVX34STYR8W7,78W56bcdQaeZ,ZaehiGPjJgkf,fgkDMlpqLVXU,123456789ABC,17DEFGHIJBKL,28MNOPHIQAKR,\break bSlmnFOijdoC,cTpENhmn9qYo.} {\tt 1}=(0,0,1,1,1,1,0,0,0,0,0,0); {\tt 2}=(0,0,1,-1,1,-1,0,0,0,0,0,0); {\tt 3}=(0,0,0,1,0,-1,0,0,0,0,0,0); {\tt 4}=(0,0,1,0,-1,0,0,0,0,0,0,0); {\tt 5}=(0,1,0,0,0,0,0,0,0,0,0,0); {\tt 6}=(1,0,0,0,0,0,0,0,0,0,0,0); {\tt 7}=(0,0,0,0,0,0,0,1,0,0,0,0); {\tt 8}=(0,0,0,0,0,0,1,0,0,0,0,0); {\tt Z}=(0,0,0,1,0,0,0,0,0,0,0,0); {\tt a}=(0,0,1,0,0,0,0,0,0,0,0,0); {\tt b}=(0,0,0,0,0,1,0,0,0,0,0,0); {\tt c}=(0,0,0,0,1,0,0,0,0,0,0,0); {\tt S}=(1,-1,1,0,1,0,0,0,0,0,0,0); {\tt T}=(1,1,0,1,0,1,0,0,0,0,0,0); {\tt U}=(1,1,0,-1,0,-1,0,0,0,0,0,0); {\tt V}=(-1,1,1,0,1,0,0,0,0,0,0,0); {\tt D}=(0,1,-1,1,0,0,1,0,0,0,0,0); {\tt M}=(1,0,1,1,0,0,0,-1,0,0,0,0); {\tt f}=(1,0,0,0,1,1,0,1,0,0,0,0); {\tt g}=(0,1,0,0,-1,1,-1,0,0,0,0,0); {\tt l}=(0,0,1,0,-1,0,1,1,0,0,0,0); {\tt p}=(0,0,0,1,0,-1,-1,1,0,0,0,0); {\tt E}=(1,0,1,0,0,-1,1,0,0,0,0,0); {\tt N}=(0,-1,1,0,0,1,0,1,0,0,0,0); {\tt h}=(-1,1,0,0,0,0,1,1,0,0,0,0); {\tt m}=(1,0,-1,-1,0,0,0,1,0,0,0,0); {\tt n}=(0,1,1,-1,0,0,-1,0,0,0,0,0); {\tt F}=(1,0,0,1,-1,0,-1,0,0,0,0,0); {\tt O}=(0,1,0,1,1,0,0,1,0,0,0,0); {\tt i}=(1,1,0,0,0,0,1,-1,0,0,0,0); {\tt G}=(0,1,0,0,1,-1,-1,0,0,0,0,0); {\tt P}=(1,0,0,0,-1,-1,0,1,0,0,0,0); {\tt H}=(1,1,0,-1,0,1,0,0,0,0,0,0); {\tt I}=(1,-1,-1,0,1,0,0,0,0,0,0,0); {\tt j}=(0,0,0,0,0,0,0,0,1,0,0,0); {\tt d}=(0,0,0,0,0,0,0,0,0,1,0,0); {\tt J}=(0,0,0,0,0,0,0,0,0,1,1,0); {\tt k}=(0,0,0,0,0,0,0,0,0,1,-1,0); {\tt W}=(0,0,0,0,0,0,0,0,1,0,1,0); {\tt Q}=(0,0,0,0,0,0,0,0,1,0,-1,0); {\tt 9}=(0,0,0,0,0,0,0,0,1,-1,0,0); {\tt e}=(0,0,0,0,0,0,0,0,0,0,0,1); {\tt A}=(0,0,0,0,0,0,0,0,1,1,1,1); {\tt B}=(0,0,0,0,0,0,0,0,1,1,-1,-1); {\tt K}=(0,0,0,0,0,0,0,0,1,-1,1,-1); {\tt X}=(0,0,0,0,0,0,0,0,1,-1,-1,-1); {\tt q}=(0,0,0,0,0,0,0,0,1,1,1,-1);\break {\tt Y}=(0,0,0,0,0,0,0,0,1,1,-1,1); {\tt L}=(0,0,0,0,0,0,0,0,1,0,0,1); {\tt o}=(0,0,0,0,0,0,0,0,0,0,1,1); {\tt C}=(0,0,0,0,0,0,0,0,0,0,1,-1); {\tt R}=(0,0,0,0,0,0,0,0,0,1,0,-1)

\subsection{\label{subsec:13ds}13D MMPHs;  16 hyperedges}
\vskip-10pt
{\bf 63-16} {\tt 123456789ABCD,123456789EFGH,12345678IJKLM,17NOPQRSTUVWM,28XYZaRS9EbBM,3478cdef9bghH,\break 3478cdef9WhiC,3478cdefIjVkM,5678lmno9LpGq,5678lmnoATrBM,lmstuvQa9WpFD,lmstuvQaEKwxM,\break ncyz!PZv9AkLM,od"OYuz!9kgiq,od"OYuz!bUjxM,efNXsty"rJwWM.} {\tt 1}=(0,0,1,1,1,1,0,0,0,0,0,0,0);\break  {\tt 2}=(0,0,1,-1,1,-1,0,0,0,0,0,0,0); {\tt 3}=(0,0,0,1,0,-1,0,0,0,0,0,0,0); {\tt 4}=(0,0,1,0,-1,0,0,0,0,0,0,0,0); {\tt 5}=(0,1,0,0,0,0,0,0,0,0,0,0,0); {\tt 6}=(1,0,0,0,0,0,0,0,0,0,0,0,0); {\tt 7}=(0,0,0,0,0,0,0,1,0,0,0,0,0); {\tt 8}=(0,0,0,0,0,0,1,0,0,0,0,0,0); {\tt l}=(0,0,0,1,0,0,0,0,0,0,0,0,0); {\tt m}=(0,0,1,0,0,0,0,0,0,0,0,0,0); {\tt n}=(0,0,0,0,0,1,0,0,0,0,0,0,0); {\tt o}=(0,0,0,0,1,0,0,0,0,0,0,0,0); {\tt c}=(1,-1,1,0,1,0,0,0,0,0,0,0,0); {\tt d}=(1,1,0,1,0,1,0,0,0,0,0,0,0); {\tt e}=(1,1,0,-1,0,-1,0,0,0,0,0,0,0); {\tt f}=(-1,1,1,0,1,0,0,0,0,0,0,0,0); {\tt N}=(0,1,-1,1,0,0,1,0,0,0,0,0,0); {\tt X}=(1,0,1,1,0,0,0,-1,0,0,0,0,0); {\tt s}=(1,0,0,0,1,1,0,1,0,0,0,0,0); {\tt t}=(0,1,0,0,-1,1,-1,0,0,0,0,0,0); {\tt y}=(0,0,1,0,-1,0,1,1,0,0,0,0,0);\break  {\tt "}=(0,0,0,1,0,-1,-1,1,0,0,0,0,0);$\>${\tt O}=(1,0,1,0,0,-1,1,0,0,0,0,0,0);$\>${\tt Y}=(0,-1,1,0,0,1,0,1,0,0,0,0,0);$\>${\tt u}=(-1,1,0,0,0,0,1,1,0,0,0,0,0);\break  {\tt z}=(1,0,-1,-1,0,0,0,1,0,0,0,0,0); {\tt !}=(0,1,1,-1,0,0,-1,0,0,0,0,0,0); {\tt P}=(1,0,0,1,-1,0,-1,0,0,0,0,0,0); {\tt Z}=(0,1,0,1,1,0,0,1,0,0,0,0,0);$\>${\tt v}=(1,1,0,0,0,0,1,-1,0,0,0,0,0);$\>${\tt Q}=(0,1,0,0,1,-1,-1,0,0,0,0,0,0);$\>${\tt a}=(1,0,0,0,-1,-1,0,1,0,0,0,0,0);\break  {\tt R}=(1,1,0,-1,0,1,0,0,0,0,0,0,0); {\tt S}=(1,-1,-1,0,1,0,0,0,0,0,0,0,0); {\tt 9}=(0,0,0,0,0,0,0,0,1,0,0,0,0); {\tt A}=(0,0,0,0,0,0,0,0,0,1,0,0,0); {\tt E}=(0,0,0,0,0,0,0,0,0,1,1,0,0); {\tt b}=(0,0,0,0,0,0,0,0,0,1,-1,0,0); {\tt T}=(0,0,0,0,0,0,0,0,1,0,1,0,0); {\tt r}=(0,0,0,0,0,0,0,0,1,0,-1,0,0);\break {\tt I}=(0,0,0,0,0,0,0,0,1,-1,0,0,0); {\tt B}=(0,0,0,0,0,0,0,0,0,0,0,1,0); {\tt J}=(0,0,0,0,0,0,0,0,1,1,1,1,0);\break {\tt K}=(0,0,0,0,0,0,0,0,1,1,-1,-1,0); {\tt w}=(0,0,0,0,0,0,0,0,1,-1,1,-1,0); {\tt U}=(0,0,0,0,0,0,0,0,1,-1,-1,-1,0);\break  {\tt j}=(0,0,0,0,0,0,0,0,1,1,1,-1,0);{\tt V}=(0,0,0,0,0,0,0,0,1,1,-1,1,0); {\tt x}=(0,0,0,0,0,0,0,0,1,0,0,1,0); {\tt k}=(0,0,0,0,0,0,0,0,0,0,1,1,0); {\tt L}=(0,0,0,0,0,0,0,0,0,0,1,-1,0); {\tt W}=(0,0,0,0,0,0,0,0,0,1,0,-1,0); {\tt M}=(0,0,0,0,0,0,0,0,0,0,0,0,1); {\tt p}=(0,0,0,0,0,0,0,0,0,1,1,1,1); {\tt F}=(0,0,0,0,0,0,0,0,0,1,-1,1,-1);$\>${\tt G}=(0,0,0,0,0,0,0,0,0,1,-1,-1,1);$\>${\tt g}=(0,0,0,0,0,0,0,0,0,1,1,-1,1);$\>${\tt h}=(0,0,0,0,0,0,0,0,0,1,1,1,-1);\break {\tt i}=(0,0,0,0,0,0,0,0,0,1,-1,1,1); {\tt H}=(0,0,0,0,0,0,0,0,0,0,0,1,1); {\tt C}=(0,0,0,0,0,0,0,0,0,0,1,0,1); {\tt D}=(0,0,0,0,0,0,0,0,0,0,1,0,-1); {\tt q}=(0,0,0,0,0,0,0,0,0,1,0,0,-1)

\subsection{\label{subsec:14ds}14D MMPH; 15 hyperedges}
\vskip-10pt
{\bf 66-15} {\tt 347E25689ABCD1,1IKLMNOPFQRSTJ,Jr\$Vx!"s\%abcke,ekdfghijlm47E3,12345679ABFGDH,27UVWXMNYZTabc,\break 347defg9ABPHnm,347defgFijkGDH,567opqrOBPFsZt,567opqr9ABaunl,567opqrijbkCGu,opvwxyLX9iQYab,\break qdz!"KWyAt\#Sab,fgIUvwz\$Oh89AB,fgIUvwz\$\#R\%jab.} {\tt 1}=(0,0,1,1,1,1,0,0,0,0,0,0,0,0);\break  {\tt 2}=(0,0,1,-1,1,-1,0,0,0,0,0,0,0,0); {\tt 3}=(0,0,0,1,0,-1,0,0,0,0,0,0,0,0); {\tt 4}=(0,0,1,0,-1,0,0,0,0,0,0,0,0,0); {\tt 5}=(0,1,0,0,0,0,0,0,0,0,0,0,0,0); {\tt 6}=(1,0,0,0,0,0,0,0,0,0,0,0,0,0); {\tt 7}=(0,0,0,0,0,0,1,0,0,0,0,0,0,0); {\tt o}=(0,0,0,1,0,0,0,0,0,0,0,0,0,0); {\tt p}=(0,0,1,0,0,0,0,0,0,0,0,0,0,0); {\tt q}=(0,0,0,0,0,1,0,0,0,0,0,0,0,0); {\tt r}=(0,0,0,0,1,0,0,0,0,0,0,0,0,0); {\tt d}=(1,-1,1,0,1,0,0,0,0,0,0,0,0,0); {\tt e}=(1,1,0,1,0,1,0,0,0,0,0,0,0,0);\break  {\tt f}=(1,1,0,-1,0,-1,0,0,0,0,0,0,0,0); {\tt g}=(-1,1,1,0,1,0,0,0,0,0,0,0,0,0); {\tt I}=(0,1,-1,1,0,0,1,0,0,0,0,0,0,0);\break  {\tt U}=(1,0,1,1,0,0,0,-1,0,0,0,0,0,0); {\tt v}=(1,0,0,0,1,1,0,1,0,0,0,0,0,0); {\tt w}=(0,1,0,0,-1,1,-1,0,0,0,0,0,0,0); {\tt z}=(0,0,1,0,-1,0,1,1,0,0,0,0,0,0); {\tt \$}=(0,0,0,1,0,-1,-1,1,0,0,0,0,0,0); {\tt J}=(1,0,1,0,0,-1,1,0,0,0,0,0,0,0);\break {\tt V}=(0,-1,1,0,0,1,0,1,0,0,0,0,0,0); {\tt x}=(-1,1,0,0,0,0,1,1,0,0,0,0,0,0); {\tt !}=(1,0,-1,-1,0,0,0,1,0,0,0,0,0,0);\break  {\tt "}=(0,1,1,-1,0,0,-1,0,0,0,0,0,0,0); {\tt K}=(1,0,0,1,-1,0,-1,0,0,0,0,0,0,0); {\tt W}=(0,1,0,1,1,0,0,1,0,0,0,0,0,0);\break  {\tt y}=(1,1,0,0,0,0,1,-1,0,0,0,0,0,0); {\tt L}=(0,1,0,0,1,-1,-1,0,0,0,0,0,0,0); {\tt X}=(1,0,0,0,-1,-1,0,1,0,0,0,0,0,0);\break  {\tt M}=(1,1,0,-1,0,1,0,0,0,0,0,0,0,0); {\tt N}=(1,-1,-1,0,1,0,0,0,0,0,0,0,0,0); {\tt O}=(0,0,0,0,0,0,0,0,0,0,1,0,0,0); {\tt h}=(0,0,0,0,0,0,0,0,1,-1,0,0,0,0); {\tt 8}=(0,0,0,0,0,0,0,0,1,1,0,0,0,0); {\tt 9}=(0,0,0,0,0,0,0,0,0,0,0,0,0,1); {\tt A}=(0,0,0,0,0,0,0,0,0,0,0,1,1,0); {\tt B}=(0,0,0,0,0,0,0,0,0,0,0,1,-1,0); {\tt P}=(0,0,0,0,0,0,0,1,0,-1,0,0,0,0); {\tt F}=(0,0,0,0,0,0,0,1,0,1,0,0,0,0); {\tt i}=(0,0,0,0,0,0,0,0,0,0,0,1,0,0); {\tt Q}=(0,0,0,0,0,0,0,0,1,0,0,0,-1,0); {\tt Y}=(0,0,0,0,0,0,0,0,1,0,0,0,1,0); {\tt s}=(0,0,0,0,0,0,0,0,1,0,0,1,1,-1); {\tt Z}=(0,0,0,0,0,0,0,0,1,0,0,-1,-1,-1);\break  {\tt t}=(0,0,0,0,0,0,0,0,1,0,0,0,0,1); {\tt \#}=(0,0,0,0,0,0,0,0,1,0,0,1,-1,-1); {\tt R}=(0,0,0,0,0,0,0,0,1,0,0,1,1,1);\break  {\tt \%}=(0,0,0,0,0,0,0,0,1,0,0,-1,0,0); {\tt j}=(0,0,0,0,0,0,0,0,0,0,0,0,1,-1); {\tt S}=(0,0,0,0,0,0,0,0,1,0,0,-1,1,-1);\break  {\tt T}=(0,0,0,0,0,0,0,0,0,0,0,1,0,-1); {\tt a}=(0,0,0,0,0,0,0,0,0,1,1,0,0,0); {\tt b}=(0,0,0,0,0,0,0,0,0,1,-1,0,0,0); {\tt c}=(0,0,0,0,0,0,0,0,1,0,0,1,-1,1); {\tt k}=(0,0,0,0,0,0,0,0,0,0,0,0,1,1); {\tt C}=(0,0,0,0,0,0,0,1,-1,1,1,0,0,0);\break  {\tt G}=(0,0,0,0,0,0,0,1,-1,-1,-1,0,0,0); {\tt u}=(0,0,0,0,0,0,0,1,1,0,0,0,0,0); {\tt D}=(0,0,0,0,0,0,0,1,1,-1,1,0,0,0);\break  {\tt E}=(0,0,0,0,0,0,0,1,0,0,-1,0,0,0); {\tt H}=(0,0,0,0,0,0,0,0,1,0,-1,0,0,0); {\tt n}=(0,0,0,0,0,0,0,1,-1,1,-1,0,0,0);\break  {\tt l}=(0,0,0,0,0,0,0,1,-1,-1,1,0,0,0);\qquad\qquad\qquad {\tt m}=(0,0,0,0,0,0,0,1,1,1,1,0,0,0)

\subsection{\label{subsec:15ds}15D MMPH; 14 hyperedges}

{\bf 66-14} {\tt fg1256v!"\#47pq3,347pqDEFlmnorsC,CAKPQRSYZabcdeT,Tcde89IJOUVhigf,1234567Ydnty"\$\%,\break 1GMRUWXYdnty"\$\%,27HNSVWXcdefghi,347CDEFhjrwxyz!,347CDEFikstuvwx,56789ABZeouz\#\$\%,56789ABabdejklm,\break 89IJOTUVYZabcde,BDLMNOPQYZabcde,EFGHIJKLYZabcde.} {\tt 1}=(0,0,1,1,1,1.0,0,0,0,0,0,0,0,0);\break  {\tt 2}=(0,0,1,-1,1,-1.0,0,0,0,0,0,0,0,0); {\tt 3}=(0,0,0,1,0,-1.0,0,0,0,0,0,0,0,0); {\tt 4}=(0,0,1,0,-1,0.0,0,0,0,0,0,0,0,0); {\tt 5}=(0,1,0,0,0,0.0,0,0,0,0,0,0,0,0); {\tt 6}=(1,0,0,0,0,0.0,0,0,0,0,0,0,0,0); {\tt 7}=(0,0,0,0,0,0.1,0,0,0,0,0,0,0,0); {\tt 8}=(0,0,0,1,0,0.0,0,0,0,0,0,0,0,0); {\tt 9}=(0,0,1,0,0,0.0,0,0,0,0,0,0,0,0); {\tt A}=(0,0,0,0,0,1.0,0,0,0,0,0,0,0,0); {\tt B}=(0,0,0,0,1,0.0,0,0,0,0,0,0,0,0); {\tt C}=(1,-1,1,0,1,0.0,0,0,0,0,0,0,0,0); {\tt D}=(1,1,0,1,0,1.0,0,0,0,0,0,0,0,0);\break  {\tt E}=(1,1,0,-1,0,-1.0,0,0,0,0,0,0,0,0); {\tt F}=(-1,1,1,0,1,0.0,0,0,0,0,0,0,0,0); {\tt G}=(0,1,-1,1,0,0.1,0,0,0,0,0,0,0,0); {\tt H}=(1,0,1,1,0,0.0,-1,0,0,0,0,0,0,0); {\tt I}=(1,0,0,0,1,1.0,1,0,0,0,0,0,0,0); {\tt J}=(0,1,0,0,-1,1.-1,0,0,0,0,0,0,0,0); {\tt K}=(0,0,1,0,-1,0.1,1,0,0,0,0,0,0,0); {\tt L}=(0,0,0,1,0,-1.-1,1,0,0,0,0,0,0,0); {\tt M}=(1,0,1,0,0,-1.1,0,0,0,0,0,0,0,0);\break  {\tt N}=(0,-1,1,0,0,1.0,1,0,0,0,0,0,0,0); {\tt O}=(-1,1,0,0,0,0.1,1,0,0,0,0,0,0,0); {\tt P}=(1,0,-1,-1,0,0.0,1,0,0,0,0,0,0,0); {\tt Q}=(0,1,1,-1,0,0.-1,0,0,0,0,0,0,0,0); {\tt R}=(1,0,0,1,-1,0.-1,0,0,0,0,0,0,0,0); {\tt S}=(0,1,0,1,1,0.0,1,0,0,0,0,0,0,0); {\tt T}=(1,1,0,0,0,0.1,-1,0,0,0,0,0,0,0); {\tt U}=(0,1,0,0,1,-1.-1,0,0,0,0,0,0,0,0); {\tt V}=(1,0,0,0,-1,-1.0,1,0,0,0,0,0,0,0);\break  {\tt W}=(1,1,0,-1,0,1.0,0,0,0,0,0,0,0,0); {\tt X}=(1,-1,-1,0,1,0.0,0,0,0,0,0,0,0,0); {\tt Y}=(0,0,0,0,0,0.0,0,0,1,1,1,1,0,0); {\tt Z}=(0,0,0,0,0,0.0,0,0,1,-1,1,-1,0,0); {\tt a}=(0,0,0,0,0,0.0,0,0,0,1,0,-1,0,0); {\tt b}=(0,0,0,0,0,0.0,0,0,1,0,-1,0,0,0); {\tt c}=(0,0,0,0,0,0.0,0,1,0,0,0,0,0,0); {\tt d}=(0,0,0,0,0,0.0,0,0,0,0,0,0,0,1); {\tt e}=(0,0,0,0,0,0.0,0,0,0,0,0,0,1,0); {\tt f}=(0,0,0,0,0,0.0,0,0,0,1,0,0,0,0); {\tt g}=(0,0,0,0,0,0.0,0,0,1,0,0,0,0,0); {\tt h}=(0,0,0,0,0,0.0,0,0,0,0,0,1,0,0); {\tt i}=(0,0,0,0,0,0.0,0,0,0,0,1,0,0,0); {\tt j}=(0,0,0,0,0,0.0,1,-1,1,0,1,0,0,0); {\tt k}=(0,0,0,0,0,0.0,1,1,0,1,0,1,0,0); {\tt l}=(0,0,0,0,0,0.0,1,1,0,-1,0,-1,0,0); {\tt m}=(0,0,0,0,0,0.0,1,-1,-1,0,-1,0,0,0); {\tt n}=(0,0,0,0,0,0.0,0,1,-1,1,0,0,1,0); {\tt o}=(0,0,0,0,0,0.0,1,0,1,1,0,0,0,-1); {\tt p}=(0,0,0,0,0,0.0,1,0,0,0,1,1,0,1); {\tt q}=(0,0,0,0,0,0.0,0,1,0,0,-1,1,-1,0); {\tt r}=(0,0,0,0,0,0.0,0,0,1,0,-1,0,1,1); {\tt s}=(0,0,0,0,0,0.0,0,0,0,1,0,-1,-1,1); {\tt t}=(0,0,0,0,0,0.0,1,0,1,0,0,-1,1,0); {\tt u}=(0,0,0,0,0,0.0,0,1,-1,0,0,-1,0,-1); {\tt v}=(0,0,0,0,0,0.0,1,-1,0,0,0,0,-1,-1); {\tt w}=(0,0,0,0,0,0.0,1,0,-1,-1,0,0,0,1); {\tt x}=(0,0,0,0,0,0.0,0,1,1,-1,0,0,-1,0); {\tt y}=(0,0,0,0,0,0.0,1,0,0,1,-1,0,-1,0); {\tt z}=(0,0,0,0,0,0.0,0,1,0,1,1,0,0,1); {\tt !}=(0,0,0,0,0,0.0,1,1,0,0,0,0,1,-1); {\tt "}=(0,0,0,0,0,0.0,0,1,0,0,1,-1,-1,0); {\tt \#}=(0,0,0,0,0,0.0,1,0,0,0,-1,-1,0,1); {\tt \$}=(0,0,0,0,0,0.0,1,1,0,-1,0,1,0,0);\qquad\qquad {\tt \%}=(0,0,0,0,0,0.0,1,-1,-1,0,1,0,0,0).

\subsection{\label{subsec:16ds}16D MMPH; 9 hyperedges}

{\bf 70-9} {\tt 28)3456Zaop\#\%7v1,17vHNSVehjqsyYwX,XYwIOTWcklnz"8)2,3478DEFGbijnu\#\$\&,56789ABCdkmqrtuy,\break 9AJKPUVWflmot!"\&,BDLQRSTUZbcgrs'(,CEMNOPQRfhipx\$\%',FGHIJKLMadegxz!(.}\break  {\tt 1}=(0,0,1,1,1,1,0,0,0,0,0,0,0,0,0,0); {\tt 2}=(0,0,1,-1,1,-1,0,0,0,0,0,0,0,0,0,0); {\tt 3}=(0,0,0,1,0,-1,0,0,0,0,0,0,0,0,0,0);\break  {\tt 4}=(0,0,1,0,-1,0,0,0,0,0,0,0,0,0,0,0); {\tt 5}=(0,1,0,0,0,0,0,0,0,0,0,0,0,0,0,0); {\tt 6}=(1,0,0,0,0,0,0,0,0,0,0,0,0,0,0,0); {\tt 7}=(0,0,0,0,0,0,0,1,0,0,0,0,0,0,0,0); {\tt 8}=(0,0,0,0,0,0,1,0,0,0,0,0,0,0,0,0); {\tt 9}=(0,0,0,1,0,0,0,0,0,0,0,0,0,0,0,0); {\tt A}=(0,0,1,0,0,0,0,0,0,0,0,0,0,0,0,0); {\tt B}=(0,0,0,0,0,1,0,0,0,0,0,0,0,0,0,0); {\tt C}=(0,0,0,0,1,0,0,0,0,0,0,0,0,0,0,0); {\tt D}=(1,-1,1,0,1,0,0,0,0,0,0,0,0,0,0,0); {\tt E}=(1,1,0,1,0,1,0,0,0,0,0,0,0,0,0,0); {\tt F}=(1,1,0,-1,0,-1,0,0,0,0,0,0,0,0,0,0); {\tt G}=(-1,1,1,0,1,0,0,0,0,0,0,0,0,0,0,0); {\tt H}=(0,1,-1,1,0,0,1,0,0,0,0,0,0,0,0,0); {\tt I}=(1,0,1,1,0,0,0,-1,0,0,0,0,0,0,0,0); {\tt J}=(1,0,0,0,1,1,0,1,0,0,0,0,0,0,0,0); {\tt K}=(0,1,0,0,-1,1,-1,0,0,0,0,0,0,0,0,0); {\tt L}=(0,0,1,0,-1,0,1,1,0,0,0,0,0,0,0,0);\break  {\tt M}=(0,0,0,1,0,-1,-1,1,0,0,0,0,0,0,0,0); {\tt N}=(1,0,1,0,0,-1,1,0,0,0,0,0,0,0,0,0); {\tt O}=(0,-1,1,0,0,1,0,1,0,0,0,0,0,0,0,0); {\tt P}=(-1,1,0,0,0,0,1,1,0,0,0,0,0,0,0,0); {\tt Q}=(1,0,-1,-1,0,0,0,1,0,0,0,0,0,0,0,0); {\tt R}=(0,1,1,-1,0,0,-1,0,0,0,0,0,0,0,0,0);\break  {\tt S}=(1,0,0,1,-1,0,-1,0,0,0,0,0,0,0,0,0); {\tt T}=(0,1,0,1,1,0,0,1,0,0,0,0,0,0,0,0); {\tt U}=(1,1,0,0,0,0,1,-1,0,0,0,0,0,0,0,0); {\tt V}=(0,1,0,0,1,-1,-1,0,0,0,0,0,0,0,0,0); {\tt W}=(1,0,0,0,-1,-1,0,1,0,0,0,0,0,0,0,0); {\tt X}=(1,1,0,-1,0,1,0,0,0,0,0,0,0,0,0,0); {\tt Y}=(1,-1,-1,0,1,0,0,0,0,0,0,0,0,0,0,0); {\tt Z}=(0,0,0,0,0,0,0,0,1,0,0,0,0,0,0,0); {\tt a}=(0,0,0,0,0,0,0,0,0,1,0,0,0,0,0,0); {\tt b}=(0,0,0,0,0,0,0,0,0,1,1,0,0,0,0,0); {\tt c}=(0,0,0,0,0,0,0,0,0,1,-1,0,0,0,0,0); {\tt d}=(0,0,0,0,0,0,0,0,1,0,1,0,0,0,0,0); {\tt e}=(0,0,0,0,0,0,0,0,1,0,-1,0,0,0,0,0); {\tt f}=(0,0,0,0,0,0,0,0,1,-1,0,0,0,0,0,0); {\tt g}=(0,0,0,0,0,0,0,0,0,0,0,1,0,0,0,0); {\tt h}=(0,0,0,0,0,0,0,0,1,1,1,1,0,0,0,0); {\tt i}=(0,0,0,0,0,0,0,0,1,1,-1,-1,0,0,0,0); {\tt j}=(0,0,0,0,0,0,0,0,1,-1,1,-1,0,0,0,0); {\tt k}=(0,0,0,0,0,0,0,0,1,-1,-1,-1,0,0,0,0); {\tt l}=(0,0,0,0,0,0,0,0,1,1,1,-1,0,0,0,0); {\tt m}=(0,0,0,0,0,0,0,0,1,1,-1,1,0,0,0,0); {\tt n}=(0,0,0,0,0,0,0,0,1,0,0,1,0,0,0,0); {\tt o}=(0,0,0,0,0,0,0,0,0,0,1,1,0,0,0,0); {\tt p}=(0,0,0,0,0,0,0,0,0,0,1,-1,0,0,0,0); {\tt q}=(0,0,0,0,0,0,0,0,0,1,0,-1,0,0,0,0); {\tt r}=(0,0,0,0,0,0,0,0,0,0,0,0,1,0,0,0); {\tt s}=(0,0,0,0,0,0,0,0,0,0,0,0,0,1,0,0); {\tt t}=(0,0,0,0,0,0,0,0,0,0,0,0,0,1,1,0); {\tt u}=(0,0,0,0,0,0,0,0,0,0,0,0,0,1,-1,0); {\tt v}=(0,0,0,0,0,0,0,0,0,0,0,0,1,0,1,0); {\tt w}=(0,0,0,0,0,0,0,0,0,0,0,0,1,0,-1,0); {\tt x}=(0,0,0,0,0,0,0,0,0,0,0,0,1,-1,0,0); {\tt y}=(0,0,0,0,0,0,0,0,0,0,0,0,0,0,0,1); {\tt z}=(0,0,0,0,0,0,0,0,0,0,0,0,1,1,1,1); {\tt !}=(0,0,0,0,0,0,0,0,0,0,0,0,1,1,-1,-1); {\tt "}=(0,0,0,0,0,0,0,0,0,0,0,0,1,-1,1,-1);\break  {\tt \#}=(0,0,0,0,0,0,0,0,0,0,0,0,1,-1,-1,-1); {\tt \$}=(0,0,0,0,0,0,0,0,0,0,0,0,1,1,1,-1); {\tt \%}=(0,0,0,0,0,0,0,0,0,0,0,0,1,1,-1,1); {\tt \&}=(0,0,0,0,0,0,0,0,0,0,0,0,1,0,0,1); {\tt '}=(0,0,0,0,0,0,0,0,0,0,0,0,0,0,1,1); {\tt (}=(0,0,0,0,0,0,0,0,0,0,0,0,0,0,1,-1); {\tt )}=(0,0,0,0,0,0,0,0,0,0,0,0,0,1,0,-1)

\subsection{\label{subsec:27ds}27D MMPH; 16 hyperedges}

{\bf 141-16} {\tt 123456789ABCDEFGHIJKLMNOPQR,9STUVWXYZIabcdefghRijklmnop,Zqrstuvwxhyz!"\#\$\%\&p'()*-/:;,\break xvw<=>?21\&\$\%@[\textbackslash ]BA;/:\^{}\_`\{KJ,91|S\}4uVwIA$\sim$a+1D\#d\%RJ+2i+3M-l:,92+4+5=t6+6wIB+7+8["F+9\%RK+A+B\_*O+C:,\break 9|3456+D+EYI$\sim$CDEF+F+GgR+2LMNO+H+Io,9+43456X+J+KI+7CDEFf+L+MR+ALMNOn+N+O,\break 9+5T+P5+6+Q+D+KI+8b+RE+9+S+F+MR+Bj+TN+C+U+H+O,9sT+P5+6W+J8I!b+RE+9e+LHR)j+TN+Cm+NQ,\break 9s=t6+6+Q+E7I!["F+9+S+GGR)\_*O+C+U+IP,q+V+5+WT+P5+6wy+X+8+Yb+RE+9\%'+Z+B+aj+TN+C:,\break +Vr+b|3456w+Xz+c$\sim$CDEF\%+Z(+d+2LMNO:,+e+f+bSTUVZw+g+h+cabcdh\%+i+j+dijklp:,\break +e+k<sU+P>+lw+g+m@!c+R\textbackslash +n\%+i+o\^{})k+T`+p:,+f+k+4+W\}3+l?w+h+m+7+Y+1C+n]\%+j+o+A+a+3L+p\{:.} {\tt 1}=(1,0,0,0,0,0,0,0,0,0,0,0,0,0,0,0,0,0,0,0,0,0,0,0,0,0,0), {\tt 2}=(0,1,0,0,0,0,0,0,0,0,0,0,0,0,0,0,0,0,0,0,0,0,0,0,0,0,0), {\tt 3}=(0,0,1,0,0,0,0,0,0,0,0,0,0,0,0,0,0,0,0,0,0,0,0,0,0,0,0), {\tt 4}=(1,1,1,1,0,0,0,0,0,0,0,0,0,0,0,0,0,0,0,0,0,0,0,0,0,0,0), {\tt 5}=(1,-1,1,-1,0,0,0,0,0,0,0,0,0,0,0,0,0,0,0,0,0,0,0,0,0,0,0), {\tt 6}=(1,-1,-1,1,0,0,0,0,0,0,0,0,0,0,0,0,0,0,0,0,0,0,0,0,0,0,0), {\tt 7}=(1,-1,-1,-1,0,0,0,0,0,0,0,0,0,0,0,0,0,0,0,0,0,0,0,0,0,0,0), {\tt 8}=(1,-1,1,1,0,0,0,0,0,0,0,0,0,0,0,0,0,0,0,0,0,0,0,0,0,0,0), {\tt 9}=(1,1,1,-1,0,0,0,0,0,0,0,0,0,0,0,0,0,0,0,0,0,0,0,0,0,0,0), {\tt A}=(1,1,0,0,0,0,0,0,0,0,0,0,0,0,0,0,0,0,0,0,0,0,0,0,0,0,0), {\tt B}=(0,0,1,1,0,0,0,0,0,0,0,0,0,0,0,0,0,0,0,0,0,0,0,0,0,0,0), {\tt C}=(0,0,1,-1,0,0,0,0,0,0,0,0,0,0,0,0,0,0,0,0,0,0,0,0,0,0,0), {\tt D}=(0,1,0,1,0,0,0,0,0,0,0,0,0,0,0,0,0,0,0,0,0,0,0,0,0,0,0), {\tt E}=(0,1,0,-1,0,0,0,0,0,0,0,0,0,0,0,0,0,0,0,0,0,0,0,0,0,0,0), {\tt F}=(1,0,-1,0,0,0,0,0,0,0,0,0,0,0,0,0,0,0,0,0,0,0,0,0,0,0,0), {\tt G}=(1,0,0,-1,0,0,0,0,0,0,0,0,0,0,0,0,0,0,0,0,0,0,0,0,0,0,0), {\tt H}=(1,0,0,1,0,0,0,0,0,0,0,0,0,0,0,0,0,0,0,0,0,0,0,0,0,0,0), {\tt I}=(0,1,-1,0,0,0,0,0,0,0,0,0,0,0,0,0,0,0,0,0,0,0,0,0,0,0,0), {\tt J}=(0,0,0,0,1,0,0,0,0,0,0,0,0,0,0,0,0,0,0,0,0,0,0,0,0,0,0), {\tt K}=(0,0,0,0,0,1,0,0,0,0,0,0,0,0,0,0,0,0,0,0,0,0,0,0,0,0,0), {\tt L}=(0,0,0,0,0,0,1,0,0,0,0,0,0,0,0,0,0,0,0,0,0,0,0,0,0,0,0), {\tt M}=(0,0,0,0,1,1,1,1,0,0,0,0,0,0,0,0,0,0,0,0,0,0,0,0,0,0,0), {\tt N}=(0,0,0,0,1,-1,1,-1,0,0,0,0,0,0,0,0,0,0,0,0,0,0,0,0,0,0,0), {\tt O}=(0,0,0,0,1,-1,-1,1,0,0,0,0,0,0,0,0,0,0,0,0,0,0,0,0,0,0,0), {\tt P}=(0,0,0,0,1,-1,-1,-1,0,0,0,0,0,0,0,0,0,0,0,0,0,0,0,0,0,0,0), {\tt Q}=(0,0,0,0,1,-1,1,1,0,0,0,0,0,0,0,0,0,0,0,0,0,0,0,0,0,0,0), {\tt R}=(0,0,0,0,1,1,1,-1,0,0,0,0,0,0,0,0,0,0,0,0,0,0,0,0,0,0,0), {\tt S}=(0,0,0,0,1,1,0,0,0,0,0,0,0,0,0,0,0,0,0,0,0,0,0,0,0,0,0), {\tt T}=(0,0,0,0,0,0,1,1,0,0,0,0,0,0,0,0,0,0,0,0,0,0,0,0,0,0,0), {\tt U}=(0,0,0,0,0,0,1,-1,0,0,0,0,0,0,0,0,0,0,0,0,0,0,0,0,0,0,0), {\tt V}=(0,0,0,0,0,1,0,1,0,0,0,0,0,0,0,0,0,0,0,0,0,0,0,0,0,0,0), {\tt W}=(0,0,0,0,0,1,0,-1,0,0,0,0,0,0,0,0,0,0,0,0,0,0,0,0,0,0,0), {\tt X}=(0,0,0,0,1,0,-1,0,0,0,0,0,0,0,0,0,0,0,0,0,0,0,0,0,0,0,0), {\tt Y}=(0,0,0,0,1,0,0,-1,0,0,0,0,0,0,0,0,0,0,0,0,0,0,0,0,0,0,0), {\tt Z}=(0,0,0,0,1,0,0,1,0,0,0,0,0,0,0,0,0,0,0,0,0,0,0,0,0,0,0), {\tt a}=(0,0,0,0,0,1,-1,0,0,0,0,0,0,0,0,0,0,0,0,0,0,0,0,0,0,0,0), {\tt b}=(0,0,0,0,0,0,0,0,1,0,0,0,0,0,0,0,0,0,0,0,0,0,0,0,0,0,0), {\tt c}=(0,1,1,1,0,0,0,0,1,0,0,0,0,0,0,0,0,0,0,0,0,0,0,0,0,0,0), {\tt d}=(0,1,-1,1,0,0,0,0,-1,0,0,0,0,0,0,0,0,0,0,0,0,0,0,0,0,0,0), {\tt e}=(0,1,1,-1,0,0,0,0,-1,0,0,0,0,0,0,0,0,0,0,0,0,0,0,0,0,0,0), {\tt f}=(0,1,1,1,0,0,0,0,-1,0,0,0,0,0,0,0,0,0,0,0,0,0,0,0,0,0,0), {\tt g}=(0,1,-1,-1,0,0,0,0,-1,0,0,0,0,0,0,0,0,0,0,0,0,0,0,0,0,0,0), {\tt h}=(0,1,1,-1,0,0,0,0,1,0,0,0,0,0,0,0,0,0,0,0,0,0,0,0,0,0,0), {\tt i}=(0,1,0,0,0,0,0,0,1,0,0,0,0,0,0,0,0,0,0,0,0,0,0,0,0,0,0), {\tt j}=(0,0,1,0,0,0,0,0,-1,0,0,0,0,0,0,0,0,0,0,0,0,0,0,0,0,0,0), {\tt k}=(0,0,0,1,0,0,0,0,-1,0,0,0,0,0,0,0,0,0,0,0,0,0,0,0,0,0,0), {\tt l}=(0,0,0,1,0,0,0,0,1,0,0,0,0,0,0,0,0,0,0,0,0,0,0,0,0,0,0), {\tt m}=(0,0,0,0,0,0,0,0,0,1,0,0,0,0,0,0,0,0,0,0,0,0,0,0,0,0,0), {\tt n}=(0,0,0,0,0,0,0,0,0,0,1,0,0,0,0,0,0,0,0,0,0,0,0,0,0,0,0), {\tt o}=(0,0,0,0,0,0,0,0,0,0,0,1,0,0,0,0,0,0,0,0,0,0,0,0,0,0,0), {\tt p}=(0,0,0,0,0,0,0,0,0,1,1,1,1,0,0,0,0,0,0,0,0,0,0,0,0,0,0), {\tt q}=(0,0,0,0,0,0,0,0,0,1,-1,1,-1,0,0,0,0,0,0,0,0,0,0,0,0,0,0), {\tt r}=(0,0,0,0,0,0,0,0,0,1,-1,-1,1,0,0,0,0,0,0,0,0,0,0,0,0,0,0), {\tt s}=(0,0,0,0,0,0,0,0,0,1,-1,-1,-1,0,0,0,0,0,0,0,0,0,0,0,0,0,0), {\tt t}=(0,0,0,0,0,0,0,0,0,1,-1,1,1,0,0,0,0,0,0,0,0,0,0,0,0,0,0), {\tt u}=(0,0,0,0,0,0,0,0,0,1,1,1,-1,0,0,0,0,0,0,0,0,0,0,0,0,0,0), {\tt v}=(0,0,0,0,0,0,0,0,0,1,1,0,0,0,0,0,0,0,0,0,0,0,0,0,0,0,0), {\tt w}=(0,0,0,0,0,0,0,0,0,0,0,1,1,0,0,0,0,0,0,0,0,0,0,0,0,0,0), {\tt x}=(0,0,0,0,0,0,0,0,0,0,0,1,-1,0,0,0,0,0,0,0,0,0,0,0,0,0,0), {\tt y}=(0,0,0,0,0,0,0,0,0,0,1,0,1,0,0,0,0,0,0,0,0,0,0,0,0,0,0), {\tt z}=(0,0,0,0,0,0,0,0,0,0,1,0,-1,0,0,0,0,0,0,0,0,0,0,0,0,0,0), {\tt !}=(0,0,0,0,0,0,0,0,0,1,0,-1,0,0,0,0,0,0,0,0,0,0,0,0,0,0,0), {\tt "}=(0,0,0,0,0,0,0,0,0,1,0,0,-1,0,0,0,0,0,0,0,0,0,0,0,0,0,0), {\tt \#}=(0,0,0,0,0,0,0,0,0,1,0,0,1,0,0,0,0,0,0,0,0,0,0,0,0,0,0), {\tt \$}=(0,0,0,0,0,0,0,0,0,0,1,-1,0,0,0,0,0,0,0,0,0,0,0,0,0,0,0), {\tt \%}=(0,0,0,0,0,0,0,0,0,0,0,0,0,1,0,0,0,0,0,0,0,0,0,0,0,0,0), {\tt \&}=(0,0,0,0,0,0,0,0,0,0,0,0,0,0,1,0,0,0,0,0,0,0,0,0,0,0,0), {\tt '}=(0,0,0,0,0,0,0,0,0,0,0,0,0,0,0,1,0,0,0,0,0,0,0,0,0,0,0), {\tt (}=(0,0,0,0,0,0,0,0,0,0,0,0,0,1,1,1,1,0,0,0,0,0,0,0,0,0,0), {\tt )}=(0,0,0,0,0,0,0,0,0,0,0,0,0,1,-1,1,-1,0,0,0,0,0,0,0,0,0,0), {\tt *}=(0,0,0,0,0,0,0,0,0,0,0,0,0,1,-1,-1,1,0,0,0,0,0,0,0,0,0,0), {\tt -}=(0,0,0,0,0,0,0,0,0,0,0,0,0,1,-1,-1,-1,0,0,0,0,0,0,0,0,0,0), {\tt /}=(0,0,0,0,0,0,0,0,0,0,0,0,0,1,-1,1,1,0,0,0,0,0,0,0,0,0,0), {\tt :}=(0,0,0,0,0,0,0,0,0,0,0,0,0,1,1,1,-1,0,0,0,0,0,0,0,0,0,0), {\tt ;}=(0,0,0,0,0,0,0,0,0,0,0,0,0,1,1,0,0,0,0,0,0,0,0,0,0,0,0), {\tt <}=(0,0,0,0,0,0,0,0,0,0,0,0,0,0,0,1,1,0,0,0,0,0,0,0,0,0,0), {\tt =}=(0,0,0,0,0,0,0,0,0,0,0,0,0,0,0,1,-1,0,0,0,0,0,0,0,0,0,0), {\tt >}=(0,0,0,0,0,0,0,0,0,0,0,0,0,0,1,0,1,0,0,0,0,0,0,0,0,0,0), {\tt ?}=(0,0,0,0,0,0,0,0,0,0,0,0,0,0,1,0,-1,0,0,0,0,0,0,0,0,0,0), {\tt @}=(0,0,0,0,0,0,0,0,0,0,0,0,0,1,0,-1,0,0,0,0,0,0,0,0,0,0,0), {\tt [}=(0,0,0,0,0,0,0,0,0,0,0,0,0,1,0,0,-1,0,0,0,0,0,0,0,0,0,0), {\tt \textbackslash}=(0,0,0,0,0,0,0,0,0,0,0,0,0,1,0,0,1,0,0,0,0,0,0,0,0,0,0), {\tt ]}=(0,0,0,0,0,0,0,0,0,0,0,0,0,0,1,-1,0,0,0,0,0,0,0,0,0,0,0), {\tt \^{}}=(0,0,0,0,0,0,0,0,0,0,0,0,0,0,0,0,0,1,0,0,0,0,0,0,0,0,0), {\tt \_}=(0,0,0,0,0,0,0,0,0,0,1,1,1,0,0,0,0,1,0,0,0,0,0,0,0,0,0), {\tt `}=(0,0,0,0,0,0,0,0,0,0,1,-1,1,0,0,0,0,-1,0,0,0,0,0,0,0,0,0), {\tt \{}=(0,0,0,0,0,0,0,0,0,0,1,1,-1,0,0,0,0,-1,0,0,0,0,0,0,0,0,0), {\tt |}=(0,0,0,0,0,0,0,0,0,0,1,1,1,0,0,0,0,-1,0,0,0,0,0,0,0,0,0), {\tt \}}=(0,0,0,0,0,0,0,0,0,0,1,-1,-1,0,0,0,0,-1,0,0,0,0,0,0,0,0,0), {\tt $\sim$}=(0,0,0,0,0,0,0,0,0,0,1,1,-1,0,0,0,0,1,0,0,0,0,0,0,0,0,0), {\tt +1}=(0,0,0,0,0,0,0,0,0,0,1,0,0,0,0,0,0,1,0,0,0,0,0,0,0,0,0), {\tt +2}=(0,0,0,0,0,0,0,0,0,0,0,1,0,0,0,0,0,-1,0,0,0,0,0,0,0,0,0), {\tt +3}=(0,0,0,0,0,0,0,0,0,0,0,0,1,0,0,0,0,-1,0,0,0,0,0,0,0,0,0), {\tt +4}=(0,0,0,0,0,0,0,0,0,0,0,0,1,0,0,0,0,1,0,0,0,0,0,0,0,0,0), {\tt +5}=(0,0,0,0,0,0,0,0,0,0,0,0,0,0,0,0,0,0,1,0,0,0,0,0,0,0,0), {\tt +6}=(0,0,0,0,0,0,0,0,0,0,0,0,0,0,0,0,0,0,0,1,0,0,0,0,0,0,0), {\tt +7}=(0,0,0,0,0,0,0,0,0,0,0,0,0,0,0,0,0,0,0,0,1,0,0,0,0,0,0), {\tt +8}=(0,0,0,0,0,0,0,0,0,0,0,0,0,0,0,0,0,0,1,1,1,1,0,0,0,0,0), {\tt +9}=(0,0,0,0,0,0,0,0,0,0,0,0,0,0,0,0,0,0,1,-1,1,-1,0,0,0,0,0), {\tt +A}=(0,0,0,0,0,0,0,0,0,0,0,0,0,0,0,0,0,0,1,-1,-1,1,0,0,0,0,0), {\tt +B}=(0,0,0,0,0,0,0,0,0,0,0,0,0,0,0,0,0,0,1,-1,-1,-1,0,0,0,0,0), {\tt +C}=(0,0,0,0,0,0,0,0,0,0,0,0,0,0,0,0,0,0,1,-1,1,1,0,0,0,0,0), {\tt +D}=(0,0,0,0,0,0,0,0,0,0,0,0,0,0,0,0,0,0,1,1,1,-1,0,0,0,0,0), {\tt +E}=(0,0,0,0,0,0,0,0,0,0,0,0,0,0,0,0,0,0,1,1,0,0,0,0,0,0,0), {\tt +F}=(0,0,0,0,0,0,0,0,0,0,0,0,0,0,0,0,0,0,0,0,1,1,0,0,0,0,0), {\tt +G}=(0,0,0,0,0,0,0,0,0,0,0,0,0,0,0,0,0,0,0,0,1,-1,0,0,0,0,0), {\tt +H}=(0,0,0,0,0,0,0,0,0,0,0,0,0,0,0,0,0,0,0,1,0,1,0,0,0,0,0), {\tt +I}=(0,0,0,0,0,0,0,0,0,0,0,0,0,0,0,0,0,0,0,1,0,-1,0,0,0,0,0), {\tt +J}=(0,0,0,0,0,0,0,0,0,0,0,0,0,0,0,0,0,0,1,0,-1,0,0,0,0,0,0), {\tt +K}=(0,0,0,0,0,0,0,0,0,0,0,0,0,0,0,0,0,0,1,0,0,-1,0,0,0,0,0), {\tt +L}=(0,0,0,0,0,0,0,0,0,0,0,0,0,0,0,0,0,0,1,0,0,1,0,0,0,0,0), {\tt +M}=(0,0,0,0,0,0,0,0,0,0,0,0,0,0,0,0,0,0,0,1,-1,0,0,0,0,0,0), {\tt +N}=(0,0,0,0,0,0,0,0,0,0,0,0,0,0,0,0,0,0,0,0,0,0,1,0,0,0,0), {\tt +O}=(0,0,0,0,0,0,0,0,0,0,0,0,0,0,0,0,0,0,0,0,0,0,0,1,0,0,0), {\tt +P}=(0,0,0,0,0,0,0,0,0,0,0,0,0,0,0,0,0,0,0,0,0,0,0,0,1,0,0), {\tt +Q}=(0,0,0,0,0,0,0,0,0,0,0,0,0,0,0,0,0,0,0,0,0,0,1,1,1,1,0), {\tt +R}=(0,0,0,0,0,0,0,0,0,0,0,0,0,0,0,0,0,0,0,0,0,0,1,-1,1,-1,0), {\tt +S}=(0,0,0,0,0,0,0,0,0,0,0,0,0,0,0,0,0,0,0,0,0,0,1,-1,-1,1,0), {\tt +T}=(0,0,0,0,0,0,0,0,0,0,0,0,0,0,0,0,0,0,0,0,0,0,1,-1,-1,-1,0), {\tt +U}=(0,0,0,0,0,0,0,0,0,0,0,0,0,0,0,0,0,0,0,0,0,0,1,-1,1,1,0), {\tt +V}=(0,0,0,0,0,0,0,0,0,0,0,0,0,0,0,0,0,0,0,0,0,0,1,1,1,-1,0), {\tt +W}=(0,0,0,0,0,0,0,0,0,0,0,0,0,0,0,0,0,0,0,0,0,0,1,1,0,0,0), {\tt +X}=(0,0,0,0,0,0,0,0,0,0,0,0,0,0,0,0,0,0,0,0,0,0,0,0,1,1,0), {\tt +Y}=(0,0,0,0,0,0,0,0,0,0,0,0,0,0,0,0,0,0,0,0,0,0,0,0,1,-1,0), {\tt +Z}=(0,0,0,0,0,0,0,0,0,0,0,0,0,0,0,0,0,0,0,0,0,0,0,1,0,1,0), {\tt +a}=(0,0,0,0,0,0,0,0,0,0,0,0,0,0,0,0,0,0,0,0,0,0,0,1,0,-1,0), {\tt +b}=(0,0,0,0,0,0,0,0,0,0,0,0,0,0,0,0,0,0,0,0,0,0,1,0,-1,0,0), {\tt +c}=(0,0,0,0,0,0,0,0,0,0,0,0,0,0,0,0,0,0,0,0,0,0,1,0,0,-1,0), {\tt +d}=(0,0,0,0,0,0,0,0,0,0,0,0,0,0,0,0,0,0,0,0,0,0,1,0,0,1,0), {\tt +e}=(0,0,0,0,0,0,0,0,0,0,0,0,0,0,0,0,0,0,0,0,0,0,0,1,-1,0,0), {\tt +f}=(0,0,0,0,0,0,0,0,0,0,0,0,0,0,0,0,0,0,0,0,0,0,0,0,0,0,1), {\tt +g}=(0,0,0,0,0,0,0,0,0,0,0,0,0,0,0,0,0,0,0,1,1,1,0,0,0,0,1), {\tt +h}=(0,0,0,0,0,0,0,0,0,0,0,0,0,0,0,0,0,0,0,1,-1,1,0,0,0,0,-1), {\tt +i}=(0,0,0,0,0,0,0,0,0,0,0,0,0,0,0,0,0,0,0,1,1,-1,0,0,0,0,-1), {\tt +j}=(0,0,0,0,0,0,0,0,0,0,0,0,0,0,0,0,0,0,0,1,1,1,0,0,0,0,-1), {\tt +k}=(0,0,0,0,0,0,0,0,0,0,0,0,0,0,0,0,0,0,0,1,-1,-1,0,0,0,0,-1), {\tt +l}=(0,0,0,0,0,0,0,0,0,0,0,0,0,0,0,0,0,0,0,1,1,-1,0,0,0,0,1), {\tt +m}=(0,0,0,0,0,0,0,0,0,0,0,0,0,0,0,0,0,0,0,1,0,0,0,0,0,0,1), {\tt +n}=(0,0,0,0,0,0,0,0,0,0,0,0,0,0,0,0,0,0,0,0,1,0,0,0,0,0,-1), {\tt +o}=(0,0,0,0,0,0,0,0,0,0,0,0,0,0,0,0,0,0,0,0,0,1,0,0,0,0,-1), {\tt +p}=(0,0,0,0,0,0,0,0,0,0,0,0,0,0,0,0,0,0,0,0,0,1,0,0,0,0,1)

\end{widetext}

%


\end{document}